\documentclass[superscriptaddress,preprint,prb,showpacs,preprintnumbers,amsmath,amssymb]{revtex4}

\usepackage [dvips]{epsfig}
\usepackage {graphics}
\usepackage {revsymb}
\usepackage{graphicx}
\usepackage{dcolumn}
\usepackage{bm}

\begin{document}


\title{Non-linear optical susceptibilities, Raman efficiencies and electrooptic
tensors from first-principles density functional perturbation theory}

\author{M. Veithen}
\affiliation{D\'epartement de Physique, Universit\'e de Li\`ege,
B-5, B-4000 Sart-Tilman, Belgium}

\author{X. Gonze}
\affiliation{Unit\'e de Physico-Chimie et de Physique des Mat\'eriaux (PCPM),
       place Croix du Sud, 1, B-1348 Louvain-la-Neuve, Belgium}

\author{Ph.Ghosez}
\affiliation{D\'epartement de Physique, Universit\'e de Li\`ege,
B-5, B-4000 Sart-Tilman, Belgium}

\date{\today}

\begin{abstract}

The non-linear response of infinite periodic solids to homogenous
electric fields and collective atomic displacements
is discussed in the framework of density functional perturbation
theory. The approach is based on the $2n + 1$ theorem
applied to an electric-field-dependent energy functional.
We report the expressions for the calculation of the non-linear optical
susceptibilities, Raman scattering efficiencies and electrooptic
coefficients. Different formulations of third-order energy derivatives are
examined and their convergence with respect to the \textbf{k}-point sampling
is discussed. We apply our method to a few simple cases and compare our
results to those obtained with distinct techniques. Finally, we discuss
the effect of a scissors correction on the EO coefficients and
non-linear optical susceptibilities.
\end{abstract}




\maketitle

\section{INTRODUCTION}

Nowadays, density functional theory~\cite{hk,ks} (DFT) is considered
as a standard method in condensed matter physics, to study electronic,
structural and macroscopic
properties of solids from first principles. Combined with adiabatic
perturbation theory, it allows {\it a priori} the computation of
derivatives of the energy and related thermodynamic potentials up to
any order. At the second
order, this approach has been applied to compute linear response functions
such as phonon frequencies or Born effective charges with an accuracy of a
few percents. The third-order derivatives are related to non-linear
properties such as phonon lifetimes, Raman tensors or non-linear optical
susceptibilities.

The linear-response formalism has been implemented in various
first-principles codes and is routinely applied to an increasing number of
systems [see for example Ref. [\onlinecite{rmp73_515}] and references
therein].
By contrast, the non-linear response formalism
has been mostly restricted to quantum chemistry problems. 
Although the hyperpolarizabilities of
a huge number of molecules have been computed, taking into account both electronic
and vibrational (ionic) contributions~\cite{irpc16_389,champagne},
applications in condensed matter physics have
focused on rather simple cases~\cite{
ssc91_813,prl75_1819,prb57_12847,prb53_15638,prb66_100301,prb67_144304,
prb69_14304,prb69_45205}.

Here, we present a methodology for the computation of nonlinear
response functions and related physical quantities of periodic
solids from density
functional perturbation theory (DFPT). We focus on perturbations
characterized by a zero wavevector and involving either three
electric fields, or two electric fields and one atomic displacement.
Following Nunes and Gonze~\cite{prb63_155107}, our approach makes use
of the $2n+1$ theorem applied to an electric-field-dependent energy
functional~\cite{prl73_712}. We report the local density approximation (LDA)
expressions, as implemented within the {\sc abinit} package~\cite{abinit}.

Our paper is organized as follows.
In Sec. \ref{sec_form}, we describe the theoretical background related to
the $2n+1$ theorem and the electric field perturbation. In Sec.
\ref{sec_nlo}, we describe the computation of the
non-linear optical susceptibilities,
the non-resonant Raman scattering efficiencies of both transverse
(TO) and longitudinal (LO) zone-center
optical phonons and the linear electrooptic (EO) tensor.
In Sec. \ref{sec_res}, we illustrate
the validity of the formalism by applying our
methodology to some semiconductors and ferroelectric oxides and 
we briefly discuss the effect of
a scissors correction on the EO coefficients and non-linear optical
susceptibilities.

Some of the tensors we consider in this work depend on static electric
fields : they include contributions of both the electrons and the
ions. Other quantities imply only the response of the valence electrons:
they are defined for frequencies of the electric field high enough to get
rid of the ionic contributions but sufficiently low to avoid electronic
excitations. For clarity, we adopt the following
convention. Static fields will be labeled by greek indices ($\alpha$,
$\beta$, ...) while we will refer to optical fields with roman symbols
($i$, $j$, ...). To simplify the notations, we will also drop
labels such as "$\infty$" for quantities that do not involve the response
of the ions. Using this convention, we can write
$\varepsilon_{ij}$ and $\varepsilon_{\alpha \beta}$,
respectively, for the optical and static dielectric tensor, respectively, and
$r_{ij \gamma}$ for the linear EO (Pockels) tensor
that involves two optical and one static
electric field.


\section{Formalism} \label{sec_form}

\subsection{Mixed third-order energy derivatives} \label{sec_der}

In this section, we present the general framework of the computation
of third order energy derivatives based on the $2n + 1$
theorem~\cite{prb39_13120,pra52_1086,pra52_1096}.
Using the notations of Refs. [\onlinecite{xggrad,xgcl}], we consider
three Hermitian perturbations labeled $\lambda_{1}$,
$\lambda_{2}$ and $\lambda_{3}$. The mixed third-order derivatives
\begin{equation} \label{eq_3dte}
E^{\lambda_1 \lambda_2 \lambda_3} =
\frac{1}{6}
\left .
\frac{\partial^3 E}
{\partial \lambda_1 \partial \lambda_{2}
\partial \lambda_{3}}
\right |_{\lambda_1 = 0, \lambda_2 = 0, \lambda_3 = 0}
\end{equation}
can be computed from the ground-state and first-order wavefunctions
\begin{widetext}
\begin{equation} \label{eq_perm}
E^{\lambda_1 \lambda_2 \lambda_3} =
\frac{1}{6} \left (
\widetilde{E}^{\lambda_1 \lambda_2 \lambda_3} +
\widetilde{E}^{\lambda_1 \lambda_3 \lambda_2} +
\widetilde{E}^{\lambda_2 \lambda_1 \lambda_3} +
\widetilde{E}^{\lambda_2 \lambda_3 \lambda_1} +
\widetilde{E}^{\lambda_3 \lambda_2 \lambda_1} +
\widetilde{E}^{\lambda_3 \lambda_1 \lambda_2}
\right )
\end{equation}
with
\begin{eqnarray} \label{eq_3dte_exp}
\widetilde{E}^{\lambda_1 \lambda_2 \lambda_3} & = &
\sum_{\alpha}
[
\langle \psi_{\alpha}^{\lambda_1} |
(T + v_{ext})^{\lambda_2 \lambda_3} | \psi_{\alpha}^{(0)} \rangle
+
\langle \psi_{\alpha}^{\lambda_1} |
(T + v_{ext} + v_{Hxc})^{\lambda_2} |
\psi_{\alpha}^{\lambda_3} \rangle \nonumber \\
& & +
\langle \psi_{\alpha}^{(0)} |
(T + v_{ext})^{\lambda_1 \lambda_2 \lambda_3} |
\psi_{\alpha}^{(0)} \rangle
+
\langle \psi_{\alpha}^{(0)} |
(T + v_{ext})^{\lambda_1 \lambda_2} | \psi_{\alpha}^{\lambda_3} \rangle
]
-
\sum_{\alpha,\beta} \Lambda_{\beta \alpha}^{\lambda_2}
\langle \psi_{\alpha}^{\lambda_1} | \psi_{\beta}^{\lambda_3} \rangle
\nonumber \\
& & +
\frac{1}{6} \int d\textbf{r} d\textbf{r}' d\textbf{r}''
\frac{\delta^3 E_{Hxc}[n^{(0)}]}
{\delta n(\textbf{r}) \delta n(\textbf{r}') \delta n(\textbf{r}'')}
n^{\lambda_1}(\textbf{r})
n^{\lambda_2}(\textbf{r}')
n^{\lambda_3}(\textbf{r}'') \nonumber \\
& & +
\frac{1}{2} \int d\textbf{r} d\textbf{r}'
\left .
\frac{d}{d \lambda_{2}}
\frac{\delta^2 E_{Hxc}[n^{(0)}]}
{\delta n(\textbf{r}) \delta n(\textbf{r}')}
\right |_{\lambda = 0}
n^{\lambda_1}(\textbf{r})
n^{\lambda_3}(\textbf{r}') \nonumber \\
& & +
\frac{1}{2} \int d\textbf{r}
\left .
\frac{d^2}{d \lambda_{1} d \lambda_{3}}
\frac{\delta E_{Hxc}[n^{(0)}]}
{\delta n(\textbf{r})}
\right |_{\lambda = 0}
n^{\lambda_2}(\textbf{r})
+
\left .
\frac{1}{6}
\frac{d^3 E_{Hxc}[n^{(0)}]}
{d \lambda_{1} d \lambda_{2} d \lambda_{3}}
\right |_{\lambda = 0}.
\end{eqnarray}
\end{widetext}
$T$ is the kinetic energy and $E_{Hxc}$ (resp. $v_{Hxc}$)
is the sum of the Hartree and
exchange-correlation energy (resp. potential).
The first-order potential $v_{Hxc}^{\lambda_2}$
can be computed as a second-order functional derivative
of $E_{Hxc}$~\cite{pra52_1096}
\begin{equation} \label{eq_vhxc1}
v_{Hxc}^{\lambda_2} = \int
\frac{\delta^2 E_{Hxc}[n^{(0)}]}{\delta n(\bm{r}) \delta n(\bm{r}')}
n^{\lambda_2}(\bm{r}') \; d \bm{r}'
+
\left .
\frac{d}{d \lambda_{2}}
\frac{\delta E_{Hxc}[n^{(0)}]}{\delta n(\bm{r})}
\right |_{\lambda = 0}.
\end{equation}
Within the parallel gauge, the first-order Lagrange multipliers are given
by
\begin{equation}
\Lambda_{\beta \alpha}^{\lambda_2}=
\langle \psi_{\beta}^{(0)} |
(T + v_{ext} + v_{Hxc})^{\lambda_2} |
\psi_{\alpha}^{(0)} \rangle.
\end{equation}
As a consequence of the "2n + 1" theorem,
the evaluation of Eq. (\ref{eq_3dte_exp}) requires no higher order derivative 
of the wavefunctions than the first one.
These first-order wavefunctions are nowadays available
in several first-principles codes. They can be computed from linear response
by minimizing a
stationary expression of the second-order energy~\cite{xggrad} or
equivalenty by solving the corresponding Sternheimer
equation~\cite{prb43_7231}.
It follows that the computation of third-order energies
does not require additional quantities than the calculation of
second-order energy derivatives.

Eq. (\ref{eq_3dte_exp}) is the general expression of the third-order
energy derivatives. In case at least one of the perturbations does not
affect the explicit form of the kinetic energy or
the Hartree and exchange-correlation energy, it can be simplified :
some of the terms may be zero. This is the case for the
electric field perturbations treated in this work as well as
for phonon type perturbations. Further simplifications can be made
in case pseudopotentials without non-linear exchange-correlation
core-correction are used.


\subsection{The electric field perturbation} \label{sec_elfd}

As mentioned in the introduction,
special care is required in case one of the perturbations $\lambda_j$
is a macroscopic electric field $\bm{\mathcal{E}}$. In fact, as 
discussed in the litterature, for
infinite periodic solids, usually treated with
Born-von Karman boundary conditions, the
scalar potential $\bm{\mathcal{E}} \cdot \textbf{r}$ breaks the periodicity of
the crystal lattice. Moreover, it is unbound from below~: it is
always possible to lower the energy by transferring electrons from the
valence states to the conduction states in a distant region (Zener
breakdown). However, for sufficiently small fields, the tunneling current
through the band gap can be neglected and the system is well described by a
set of electric field dependent Wannier functions
$W_n(\textbf{r})$.
As shown by Nunes and Vanderbilt~\cite{prl73_712},
these Wannier functions minimize the energy functional
\begin{equation} \label{eq_energy}
E \left [
W_n; \bm{\mathcal{E}}
\right ] =
E_0 \left [ W_n \right ]
-
\Omega \bm{\mathcal{E}} \cdot \bm{\mathcal{P}}
\end{equation}
where $E_0$ is the Kohn-Sham energy under zero field and $\bm{\mathcal{P}}$ the
macroscopic polarization that can be computed from the Wannier function
centers.
It is important to note that these Wannier functions do not correspond to
the true ground-state of the system but rather to a long lived metastable
state.

In practical applications, it is
not mandatory to evaluate the functional Eq. (\ref{eq_energy}) in a Wannier
basis. It can equivalently be expressed using Bloch functions
$u_{n \textbf{k}}$ related to $W_n$ by a
unitary transform. In this case, the polarization can be computed as a
Berry phase of the occupied bands~\cite{prb47_1651}
\begin{equation} \label{eq_ksvb}
\bm{\mathcal{P}} =
-\frac{2ie}{(2 \pi)^3}
\sum_{n}^{occ}
\int_{BZ} d \textbf{k}
\langle u_{n \textbf{k}} | \bm{\nabla}_k | u_{n \textbf{k}} \rangle
\end{equation}
where BZ is the Brillouin zone, $e$ the absolute value of the electronic
charge and the factor of 2 accounts for spin degeneracy.
The Bloch functions are chosen to satisfy the periodic gauge condition
\begin{equation} \label{eq_pergauge}
e^{i \textbf{G} \cdot \textbf{R}} u_{n \textbf{k}+\textbf{G}}
=
u_{n \textbf{k}}.
\end{equation}
In order to use Eq. (\ref{eq_ksvb}) in practical calculations, the integration
over the BZ, as well as the differentiation with respect to \textbf{k}, have
to be performed on a discrete mesh of $N_k$ {\bf k}-points.
In case of the ground-state polarization,
the standard approach is to build
strings of \textbf{k}-points parallel to a vector of the reciprocal
space $\textbf{G}_{\parallel}$. The polarization can then be computed as
a string-averaged Berry phase.
Unfortunately, the adaptation of this method 
to the computation of the energy derivatives is plagued
with several difficulties, like the following.
The general form of the non-linear optical susceptibility
tensor of a compound is imposed by its symmetry.
For example, in zinc-blende semiconductors, 
this tensor, expressed in {\it cartesian coordinates}
reduces to
$\chi^{(2)}_{ijl} = \chi^{(2)} |\epsilon_{ijl}|$
where $\epsilon$ is the Levi-Civita tensor.
It follows that the {\it reduced coordinates} formulation
of $\chi^{(2)}_{ijl}$ satisfy
the relation
\begin{equation} \label{eq_chizb}
\left |
\frac{\chi^{(2)}_{ijl}}{\chi^{(2)}_{iii}}
\right |
= \frac{1}{3}
\end{equation}
where at least one of the three indices $i$, $j$ and $l$
are different from the two others.
When we tried to use strings of {\bf k}-points to compute $\chi^{(2)}_{ijl}$,
Eq. (\ref{eq_chizb}) was not satisfied.
However, we were able to avoid these problems, by using
the finite difference formula of Marzari and Vanderbilt~\cite{marzari_wannier}
on a regular grid of
special \textbf{k}-points (instead of strings) 
\begin{equation} \label{eq_mv}
\bm{\nabla} f (\textbf{k}) =
\sum_{\textbf{b}} w_{\textbf{b}} \textbf{b}
\left [
f (\textbf{k}+\textbf{b}) - f (\textbf{k})
\right ]
\end{equation}
where \textbf{b} is a vector connecting a \textbf{k}-point to one of 
its nearest
neighbours and $w_{\textbf{b}}$ is a weight factor. The sum in Eq.
(\ref{eq_mv}) includes as many shells of first neighbours as necessary to
satisfy the condition
\begin{equation} \label{eq_wb}
\sum_{\textbf{b}} w_{\textbf{b}} b_{\alpha} b_{\beta}
=
\frac{g_{\alpha \beta}}{(2 \pi)^2}
\end{equation}
where $b_{\alpha}$ are the reduced coordinates of \textbf{b}
and $g_{\alpha \beta}$ is the metric tensor associated to the real space
crystal lattice.

In case of the ground-state polarization, we cannot apply the
discretization Eq. (\ref{eq_mv}) directly to Eq. (\ref{eq_ksvb}).
As shown by Marzari and Vanderbilt, the
discretization of Eq. (\ref{eq_ksvb}) does not transform correctly under
the gauge transformation
\begin{equation} \label{eq_gauge}
u_{n \bm{k}} (\bm{r}) \rightarrow
e^{-i \bm{k} \cdot \bm{R}} u_{n \bm{k}} (\bm{r}).
\end{equation}
Since Eq. (\ref{eq_gauge}) is equivalent to a shift of the origin by one
lattice vector $\bm{R}$, $\bm{\mathcal{P}}$ must
change accordingly by one polarization quantum.
In order to obtain a discrete expression that matches this
requirement, we must combine Eq. (\ref{eq_mv}) with
the King-Smith and
Vanderbilt formula~\cite{prb47_1651,prb48_4442}
\begin{equation} \label{eq_poldisc}
\bm{\mathcal{P}} =
\frac{2e}{N_k \Omega} \sum_{\textbf{k}} \sum_{\textbf{b}}
w_{\textbf{b}} \textbf{b} \Im \ln \det
\left [
S(\textbf{k},\textbf{k}+\textbf{b})
\right ]
\end{equation}
where $S$ is the overlap matrix between Bloch functions at \textbf{k} and
$\textbf{k}+\textbf{b}$:
\begin{equation} \label{eq_overlap}
S_{n,m}(\textbf{k},\textbf{k}+\textbf{b}) =
\langle u_{n \textbf{k}} | u_{m,\textbf{k}+\textbf{b}} \rangle.
\end{equation}

As discussed by Nunes and Gonze~\cite{prb63_155107}, when we
apply the perturbation expansion of the preceeding section to the energy
functional Eq. (\ref{eq_energy}), we can adopt two equivalent approaches.
The first possibility is the use of
Eq. (\ref{eq_ksvb}) for the polarization and a
discretization after having performed the perturbation expansion.
The second possibility is to apply the $2n + 1$ theorem directly to
Eq. (\ref{eq_poldisc}) in which case no additional discretization is
needed. Using the notations of Nunes and Gonze,
we will refer to the first case as the
DAPE (discretization after perturbation expansion) formulation
and to the second one as the
PEAD (perturbation expansion after discretization)
formulation of the third-order energy.
In the following sections,
we will discuss both expressions.
In addition, in Sec. \ref{sec_res_sc},
we will compare
their convergence with respect to the \textbf{k}-point sampling on a
realistic example.
The perturbation expansion of the first term
($E_0$) of Eq. (\ref{eq_energy}) can easily be performed as
it is described in the Sec. \ref{sec_der}. At the opposite,
the expansion of the second term
($-\Omega \bm{\mathcal{E}}\cdot \bm{\mathcal{P}}$)
is more tricky since it explicitely depends on the polarization.
In the two sections that follow, we will focus on
the $-\Omega \bm{\mathcal{E}}\cdot \bm{\mathcal{P}}$
term of Eq. (\ref{eq_energy}).
It will be referred to as $E_{pol}$.


\subsection{DAPE expression}

According to the formalism of the preceeding section, the
$\bm{\mathcal{E}} \cdot \bm{\mathcal{P}}$ term
acts as an additional external potential that has to be added to the ionic
one. The $\bm{\mathcal{E}} \cdot \bm{\mathcal{P}}$ perturbation
is linear in the electric field and
does not depend explicitely on other variables such as
the atomic positions. It just enters the terms of Eq. (\ref{eq_3dte_exp})
that involve the first derivative of $v_{ext}$ with respect to
$\bm{\mathcal{E}}$. In other words, the only terms in Eq. (\ref{eq_perm})
that involve the expansion of $\bm{\mathcal{P}}$ are of the form
$\widetilde{E}^{\lambda_1 \mathcal{E}_{i} \lambda_3}$
where $\lambda_1$ and $\lambda_3$
represent an arbitrary perturbation such as an electric field or an atomic
displacement.

The DAPE expression of the third-order derivative
of E$_{pol}$ writes as follows
\begin{equation} \label{eq_dape}
\widetilde{E}_{pol}^{\lambda_1 \mathcal{E}_{i} \lambda_3}
=
\frac{2ie \Omega}{(2 \pi)^3}
\int_{BZ} d \bm{k}
\sum_{n}^{occ}
\langle u_{n \bm{k}}^{\lambda_1} |
\left (
\frac{\partial}{\partial k_{i}}
\sum_{m}^{occ}
| u_{m \bm{k}}^{\lambda_3} \rangle
\langle u_{m \bm{k}}^{(0)} |
\right )
| u_{n \bm{k}}^{(0)} \rangle
\end{equation}
where $u_{n \bm{k}}^{\lambda_j}$ are the projection of the first-order
wavefunctions on the conduction bands.
The complete expression of various third-order
energy derivatives,
taking into account the expansion of both E$_0$ and E$_{pol}$,
are reported in Sec. \ref{sec_nlo}.
Eq. (\ref{eq_dape}) was derived first by Dal Corso and
Mauri~\cite{prb50_5756} in a slightly different context: they
performed the
perturbation expansion of the energy functional Eq. (\ref{eq_energy})
using a Wannier basis. The resulting expression of the third-order energy was
expressed in terms of Bloch functions by applying a unitary transform to the
Wannier orbitals.

Using the finite difference expression of Marzari and
Vanderbilt Eq. (\ref{eq_mv}), Eq. (\ref{eq_dape}) becomes
\begin{eqnarray} \label{eq_dape_mv}
\widetilde{E}_{pol}^{\lambda_1 \mathcal{E}_{i} \lambda_3}
& = &
\frac{2ie}{N_k}
\sum_{\bm{k}} \sum_{\bm{b}} \sum_{n,m}^{occ}
w_{\bm{b}} (\bm{b} \cdot \bm{G}_{i}) \nonumber \\
& &
\times
\left \{
\langle u_{n \bm{k}}^{\lambda_1} | u_{m \bm{k}+\bm{b}}^{\lambda_3} \rangle
\langle u_{m \bm{k}+\bm{b}}^{(0)} | u_{n \bm{k}}^{(0)} \rangle
\right . \nonumber \\
& & -
\left .
\langle u_{n \bm{k}}^{\lambda_1} | u_{m \bm{k}}^{\lambda_3} \rangle
\delta_{n,m}
\right \}
\end{eqnarray}
where $G_i$ is a basis vector of the reciprocal lattice.


\subsection{PEAD expression}

Applying directly the $2n + 1$ theorem
to Eq. (\ref{eq_poldisc}) we obtain
the alternative PEAD formulation of the third-order energy:
\begin{eqnarray} \label{eq_pead}
\widetilde{E}_{pol}^{\lambda_1 \mathcal{E}_{i} \lambda_3}
& = &
\frac{-e}{N_k} \Im
\sum_{\bm{k}} \sum_{\bm{b}} w_{\bm{b}} (\bm{b} \cdot \bm{G}_{i})
\nonumber \\
& &
\times
\Bigglb [
2 \sum_{n,m}^{occ}
\langle u_{n \bm{k}}^{\lambda_1} | u_{m \bm{k}+\bm{b}}^{\lambda_3} \rangle
Q_{m n}(\bm{k},\bm{k}+\bm{b})
\nonumber \\
& &
-
\sum_{n,m,l,l'}^{occ}
S^{\lambda_1}_{mn}(\bm{k},\bm{k}+\bm{b})Q_{nl}(\bm{k},\bm{k}+\bm{b})
\nonumber \\
& &
\times
S^{\lambda_3}_{ll'}(\bm{k},\bm{k}+\bm{b})Q_{l'm}(\bm{k},\bm{k}+\bm{b})
\Biggrb ]
\end{eqnarray}
where $Q$ is the inverse of the overlap matrix $S$ and $S^{\lambda_j}$ its
first-order perturbation expansion
\begin{equation} \label{eq_s1}
S^{\lambda_j}_{nm}(\bm{k},\bm{k}+\bm{b}) =
\langle u_{n \bm{k}}^{\lambda_j} | u_{m \bm{k}+\bm{b}}^{(0)} \rangle
+
\langle u_{n \bm{k}}^{(0)} | u_{m \bm{k}+\bm{b}}^{\lambda_j} \rangle.
\end{equation}


\section{Non-linear optical properties} \label{sec_nlo}

In the previous section we have discussed the general expressions
of third energy derivatives. We now particularize them to the
computation of selected non-linear properties.

\subsection{Non-linear optical susceptibilities}

In an insulator the polarization can be expressed as a Taylor
expansion of the macroscopic electric field
\begin{equation} \label{eq_nlo_pol}
P_{i} = P_{i}^s +
\sum_{j=1}^{3} \chi^{(1)}_{ij} \mathcal{E}_{j} +
\sum_{j,k=1}^{3}
\chi^{(2)}_{ijl} \mathcal{E}_{j}
   \mathcal{E}_{l} + \cdots
\end{equation}
where $P_{i}^s$ is the zero-field (spontaneous) polarization,
$\chi^{(1)}_{ij}$ the linear dielectric susceptibility
(second rank tensor) and $\chi^{(2)}_{ijl}$ the second-order non-linear
optical susceptibility (third rank tensor).
In the litterature on non-linear optics, one often finds another definition
of the non-linear optical susceptibility : instead of 
$\chi^{(2)}_{ijl}$, it is more convenient to rely on 
the $\bm{d}$-tensor defined as
\begin{equation} \label{eq_dtensor}
d_{ijl} = \frac{1}{2} \chi^{(2)}_{ijl}.
\end{equation}

In general, the polarization depends on valence electrons as well as
ions. In the present section, we deal only with the electronic 
contribution : we will consider the ionic cores as clamped to their equilibrum
positions. This constraint will be relaxed in the following
sections where we allow for
ionic displacements.

Experimentally, the electronic
contribution to the linear and non-linear susceptibilities
corresponds to measurements for electric fields at frequencies
high enough to get rid of
the ionic relaxation but low enough to avoid electronic
excitations. In case of the second-order susceptibilities,
this constraint implies that both
the frequency of $\bm{\mathcal{E}}$,
and its second harmonic, are lower than the
fundamental absorption gap.

The general expression of the electronic
non-linear optical susceptibility depends on
the frequencies of the optical electric fields [see
for example Ref. [\onlinecite{prb53_10751}]]. In the present context of the
$2n + 1$ theorem applied within the LDA to (static) DFT, we neglect the
dispersion of $\chi^{(2)}_{ijl}$.
As a consequence, $\chi^{(2)}_{ijl}$ satisfies Kleinman's~\cite{pr126_1977}
symmetry condition which means that it is symmetric under a permutation
of $i$, $j$ and $l$.
In order to be able to investigate its
frequency dependence, one would need either to apply the formalism of
time-dependent DFT~\cite{prb53_15638} or to use expressions that involve sums
over excited states~\cite{prl66_41,prb44_12781,prb49_4532,prb47_9464,prb67_165332}.
Following the work of Dal Corso and
co-workers~\cite{prb53_15638,prb50_5756} we can relate the non-linear optical
susceptibilities to a third-order derivative of the energy with respect to
an electric field
\begin{equation} \label{eq_ec_rel}
\chi^{(2)}_{ijl} = - \frac{3}{\Omega}
E^{\mathcal{E}_{i}
\mathcal{E}_{j}
\mathcal{E}_{l}}
\end{equation}
where
$
E^{\mathcal{E}_{i}
\mathcal{E}_{j}
\mathcal{E}_{l}}
$
is defined as the sum over the permutations of the three perturbations
$
\widetilde{E}^{\mathcal{E}_{i}
\mathcal{E}_{j}
\mathcal{E}_{l}}
$
(\ref{eq_perm}). Using the PEAD formulation of
Sec. \ref{sec_elfd} we can express these terms as follows
\begin{widetext}
\begin{eqnarray} \label{eq_e3_elfd}
\widetilde{E}^{\mathcal{E}_{i}
\mathcal{E}_{j}
\mathcal{E}_{l}}
& = &
\frac{-e}{N_k} \Im
\sum_{\bm{k}} \sum_{\bm{b}} w_{\bm{b}} (\bm{b} \cdot \bm{G}_{j})
\Bigglb [
2 \sum_{n,m}^{occ}
\langle u_{n \bm{k}}^{\mathcal{E}_{i}} |
u_{m \bm{k}+\bm{b}}^{\mathcal{E}_{l}} \rangle
Q_{m n}(\bm{k},\bm{k}+\bm{b})
\nonumber \\
& &
-
\sum_{n,m,n',m'}^{occ}
S^{\mathcal{E}_{i}}_{mn}(\bm{k},\bm{k}+\bm{b})
Q_{nn'}(\bm{k},\bm{k}+\bm{b})
S^{\mathcal{E}_{l}}_{n'm'}(\bm{k},\bm{k}+\bm{b})
Q_{m'm}(\bm{k},\bm{k}+\bm{b})
\Biggrb ] \nonumber \\
& &
+
\frac{2}{N_k} \sum_{\bm{k}} \sum_{n,m}^{occ}
\Bigglb [
\delta_{m,n}
\langle u_{n \textbf{k}}^{\mathcal{E}_{i}}  |
v_{hxc}^{\mathcal{E}_{j}}
| u_{m \textbf{k}}^{\mathcal{E}_{l}} \rangle
-
\langle u_{m \textbf{k}}^{(0)} |
v_{hxc}^{\mathcal{E}_{j}}
| u_{n \textbf{k}}^{(0)}  \rangle
\langle u_{n \textbf{k}}^{\mathcal{E}_{i}} |
u_{m \textbf{k}}^{\mathcal{E}_{l}}  \rangle
\Bigglb ] \nonumber \\
& & +
\frac{1}{6} \int d\textbf{r} d\textbf{r}' d\textbf{r}''
\frac{\delta^3 E_{xc}[n^0]}
{\delta n(\textbf{r}) \delta n(\textbf{r}') \delta n(\textbf{r}'')}
n^{\mathcal{E}_i}(\textbf{r})
n^{\mathcal{E}_j}(\textbf{r}')
n^{\mathcal{E}_l}(\textbf{r}'').
\end{eqnarray}
\end{widetext}


\subsection{Raman susceptibilities of zone-center optical phonons}
\label{sec_raman}

We now consider the computation of Raman scattering
efficiencies of zone-center optical phonons.
In the limit $\bm{q} \rightarrow 0$, the matrix of
interatomic force constants $\widetilde{C}$ can be expressed as the sum of
an analytical part and a non-analytical
term~\cite{xgcl}
\begin{equation} \label{eq_ifc}
\widetilde{C}_{\kappa \alpha, \kappa' \beta}
(\bm{q} \rightarrow 0)
=
\widetilde{C}_{\kappa \alpha, \kappa' \beta}^{AN}
(\bm{q} = 0)
+
\widetilde{C}_{\kappa \alpha, \kappa' \beta}^{NA}
(\bm{q} \rightarrow 0).
\end{equation}
The analytical part corresponds to the second-order derivative of the
energy with respect to an atomic displacement at $\bm{q} = 0$ under the
condition of vanishing macroscopic electric field. The second term is due
to the long-range electrostatic interactions in polar crystals. It is at
the origin of the so-called LO-TO splitting and can be computed from the
knowledge of the
Born effective charges
$
Z^{\ast}_{\kappa \alpha \beta}
$
and the electronic dielectric tensor~\cite{xgcl}
$\varepsilon_{ij}$.
The phonon frequencies $\omega_m$ and eigendisplacements
$u_m (\kappa \alpha)$ are solution of the following generalized eigenvalue
problem
\begin{equation} \label{eq_eigen}
\sum_{\kappa', \beta} \widetilde{C}_{\kappa \alpha, \kappa' \beta}
u_m (\kappa' \beta) =
M_{\kappa} \omega^2_m u_m (\kappa \alpha)
\end{equation}
where $M_{\kappa}$ is the mass of atom $\kappa$. As a convention, we choose
the eigendisplacements to be normalized as
\begin{equation} \label{eq_norm}
\sum_{\kappa, \alpha} M_{\kappa}
u_m (\kappa \alpha) u_n (\kappa \alpha)
=
\delta_{m,n}.
\end{equation}

In what follows we consider non-resonant Raman
scattering where an incoming photon of frequency $\omega_0$ and polarization
$\textbf{e}_0$ is scattered to an outgoing photon of frequency
$(\omega_0-\omega_m)$ and polarization $\textbf{e}_S$ by creating a
phonon of frequency $\omega_m$ (Stokes process).
The scattering efficiency~\cite{cardona,venkataraman} (cgs units)
corresponds to
\begin{eqnarray} \label{eq_efficiency}
\frac{d S}{d \Omega} & = &
\left |
\textbf{e}_S \cdot R^m \cdot \textbf{e}_0
\right |^2 \nonumber \\
& = &
\frac{(\omega_0 - \omega_m)^4}{c^4}
\left |
\textbf{e}_S \cdot \bm{\alpha}^m \cdot \textbf{e}_0
\right |^2
\frac{\hbar}{2 \omega_m}
(n_m + 1)
\end{eqnarray}
where $c$ is the speed of light in vacuum and $n_m$ the Boson factor
\begin{equation}
n_m = \frac{1}{\exp(\hbar \omega_m/k_B T)-1}.
\end{equation}
The Raman susceptibility $\bm{\alpha}^m$ is defined as
\begin{equation} \label{eq_ramansus}
\alpha_{ij}^m =
\sqrt{\Omega} \sum_{\kappa,\beta}
\frac{\partial \chi_{ij}^{(1)}}
{\partial \tau_{\kappa \beta}}
u_m (\kappa \beta)
\end{equation}
where $\chi_{ij}^{(1)}$ is the electronic
linear dielectric susceptibility tensor.
$\Omega$ is the angle of collection in which the outgoing photon
is scattered. Due to Snell's law, $\Omega$ is modified at the interface
between the sample and the surrounding medium. Experimentally, the
scattering efficiencies are measured with respect to the solid angle
of the medium while Eq. (\ref{eq_efficiency}) refers to the solid angle
inside the sample. In order to relate theory and experiment,
one has to take into account the
different refractive indices of the sample and medium.
For example, in the isotropic case, Eq. (\ref{eq_efficiency}) 
has to be multiplied~\cite{cardona}
by $(n'/n)^2$ where $n$ and $n'$ are respectively the refractive indices
of the sample and the medium.

For pure transverse optical phonons,
$
\frac{\partial \chi_{ij}^{(1)}}
{\partial \tau_{\kappa \beta}}
$
can be computed as a mixed third-order derivative of the energy with respect
to an electric field, twice, and to an
atomic displacement under the condition
of zero electric field
\begin{equation} \label{eq_dchidtau}
\left .
\frac{\partial \chi_{ij}^{(1)}}
{\partial \tau_{\kappa \lambda}}
\right |_{\mathcal{E}=0}
=
-\frac{6}{\Omega}
E^{\tau_{\kappa \lambda} \mathcal{E}_{i} \mathcal{E}_{j}}.
\end{equation}
In case of longitudinal optical phonons with wavevector
$\bm{q} \rightarrow 0$ in a polar crystal, Eq. (\ref{eq_ramansus})
must take into account the effect of the
macroscopic electric field generated by the lattice polar vibration. This field
enters the computation of the Raman susceptibilities at two levels.
On one hand, it
gives rise to the non-analytical part of the matrix of interatomic force
constants Eq. (\ref{eq_ifc}) that modifies the frequencies and eigenvectors with
respect to pure transverse phonons.
On the other hand, the electric field induces an additional change in the
dielectric susceptibility tensor related to the non-linear optical
coefficients $\chi^{(2)}_{ijk}$.
For longitudinal optical phonons, Eq. (\ref{eq_dchidtau}) has to be
modified as follows~\cite{prb1_3494}:
\begin{equation} \label{eq_dchidtau_long}
\frac{\partial \chi_{ij}}
{\partial \tau_{\kappa \lambda}}
=
\left .
\frac{\partial \chi_{ij}}
{\partial \tau_{\kappa \lambda}}
\right |_{\mathcal{E}=0}
-
\frac{8 \pi}{\Omega}
\frac{
\sum_{l} Z^{\ast}_{\kappa \lambda l} q_{l}
}
{\sum_{l,l'} q_{l} \varepsilon_{ll'}
q_{l'}
}
\sum_{l} \chi^{(2)}_{ijl} q_{l}.
\end{equation}

The mixed third-order derivatives (\ref{eq_dchidtau}) can be computed from
various techniques including finite differences of the dielectric
tensor~\cite{prb33_5969,prb63_94305,prl90_27401}
or the second derivative of the
electronic density matrix~\cite{prl90_36401,prb68_161101}.
Here, we follow an approach similar to Deinzer and Strauch~\cite{prb66_100301}
based on the $2n+1$ theorem. The third-order energy can be computed as the
sum over the 6 permutations Eq. (\ref{eq_perm}) of
$\tau_{\kappa \lambda}$, $\mathcal{E}_{i}$ and
$\mathcal{E}_{j}$. According to the discussion of Sec. \ref{sec_elfd},
we have to distinguish between the terms that involve the discretization of
the polarization such as
$
\widetilde{E}^{\tau_{\kappa \lambda} \mathcal{E}_{i}
\mathcal{E}_{j}}
$
or
$
\widetilde{E}^{\mathcal{E}_{j} \mathcal{E}_{i}
\tau_{\kappa \lambda}}
$
and those that can be computed from a straightforward application of the
$2n+1$ theorem such as
$
\widetilde{E}^{\mathcal{E}_{i} \tau_{\kappa \lambda}
\mathcal{E}_{j}}
$.
The former ones show an electric field as second perturbation. They can be
computed from an expression analogous to Eq. (\ref{eq_e3_elfd})
\begin{widetext}
\begin{eqnarray} \label{eq_e3_tee}
\widetilde{E}^{\tau_{\kappa \lambda}
\mathcal{E}_{i}
\mathcal{E}_{j}}
& = &
\frac{-e}{N_k} \Im
\sum_{\bm{k}} \sum_{\bm{b}} w_{\bm{b}} (\bm{b} \cdot \bm{G}_{i})
\Bigglb [
2 \sum_{n,m}^{occ}
\langle u_{n \bm{k}}^{\tau_{\kappa \lambda}} |
u_{m \bm{k}+\bm{b}}^{\mathcal{E}_{j}} \rangle
Q_{m n}(\bm{k},\bm{k}+\bm{b})
\nonumber \\
& &
-
\sum_{n,m,l,l'}^{occ}
S^{\tau_{\kappa \lambda}}_{mn}(\bm{k},\bm{k}+\bm{b})
Q_{nl}(\bm{k},\bm{k}+\bm{b})
S^{\mathcal{E}_{j}}_{ll'}(\bm{k},\bm{k}+\bm{b})
Q_{l'm}(\bm{k},\bm{k}+\bm{b})
\Biggrb ] \nonumber \\
& &
+
\frac{2}{N_k} \sum_{\bm{k}} \sum_{n,m}^{occ}
\Bigglb [
\delta_{m,n}
\langle u_{n \textbf{k}}^{\tau_{\kappa \lambda}}  |
v_{hxc}^{\mathcal{E}_{i}}
| u_{m \textbf{k}}^{\mathcal{E}_{j}} \rangle
-
\langle u_{m \textbf{k}}^{(0)} |
v_{hxc}^{\mathcal{E}_{i}}
| u_{n \textbf{k}}^{(0)}  \rangle
\langle u_{n \textbf{k}}^{\tau_{\kappa \lambda}} |
u_{m \textbf{k}}^{\mathcal{E}_{j}}  \rangle
\Bigglb ] \nonumber \\
& & +
\frac{1}{6} \int d\textbf{r} d\textbf{r}' d\textbf{r}''
\frac{\delta^3 E_{xc}[n^0]}
{\delta n(\textbf{r}) \delta n(\textbf{r}') \delta n(\textbf{r}'')}
n^{\tau_{\kappa \lambda}}(\textbf{r})
n^{\mathcal{E}_{i}}(\textbf{r}')
n^{\mathcal{E}_{j}}(\textbf{r}'').
\end{eqnarray}
We obtain a similar expression for
$
\widetilde{E}^{\mathcal{E}_{j} \mathcal{E}_{i}
\tau_{\kappa \lambda}}
$.
The remaining terms do not require any differentiation with respect to
\textbf{k}. They can be computed from the first-order change of the ionic
(pseudo) potential with respect to an atomic displacement
$v_{ext}^{\tau_{\kappa \lambda}}$
\begin{eqnarray} \label{eq_e3_ete}
\widetilde{E}^{\mathcal{E}_{i} \tau_{\kappa \lambda}
\mathcal{E}_{j}} = & &
\frac{2}{N_k} \sum_{\bm{k}} \sum_{n,m}^{occ}
\Bigglb [
\langle u_{n \textbf{k}}^{\mathcal{E}_{i}} |
v_{ext}^{\tau_{\kappa \lambda}} +
v_{hxc}^{\tau_{\kappa \lambda}} |
u_{m \textbf{k}}^{\mathcal{E}_{j}} \rangle
\delta_{n,m}
-
\langle u_{n \textbf{k}}^{(0)} |
v_{ext}^{\tau_{\kappa \lambda}} +
v_{hxc}^{\tau_{\kappa \lambda}} |
u_{m \textbf{k}}^{(0)} \rangle
\langle u_{m \textbf{k}}^{\mathcal{E}_{i}} |
u_{n \textbf{k}}^{\mathcal{E}_{j}} \rangle
\Bigglb ] \nonumber \\
& &
+ \frac{1}{2} \int d\textbf{r} d\textbf{r}'
\left .
\frac{d}{d \tau_{\kappa \lambda}}
\frac{\delta^2 E_{Hxc}}
{\delta n(\textbf{r}) \delta n(\textbf{r}')}
\right |_{n^{(0)}}
n^{\mathcal{E}_{i}}(\textbf{r})
n^{\mathcal{E}_{j}}(\textbf{r}') \nonumber \\
& &
\frac{1}{6} \int d\textbf{r} d\textbf{r}' d\textbf{r}''
\frac{\delta^3 E_{xc}[n^0]}
{\delta n(\textbf{r}) \delta n(\textbf{r}') \delta n(\textbf{r}'')}
n^{\tau_{\kappa \lambda}}(\textbf{r})
n^{\mathcal{E}_{i}}(\textbf{r}')
n^{\mathcal{E}_{j}}(\textbf{r}'').
\end{eqnarray}
\end{widetext}
In pseudopotential calculations, the computation of the first-order ionic
potential
$v_{ext}^{\tau_{\kappa \lambda}}$
requires the derivative of local and non-local (usually separable) operators.
These operations can be performed easily without any additional workload
as described in Ref. [\onlinecite{xggrad}].

In spite of the similarities between Eqs. (\ref{eq_e3_tee})
and (\ref{eq_e3_ete}) and the expression proposed by Deinzer and Strauch we
can quote few differences. First, our expression of the third-order energy
makes use of the PEAD fomulation for the expansion of the
polarization. Moreover, Eq. (\ref{eq_e3_ete}) is more general
since it allows the use of pseudopotentials with non-linear core
correction through the derivative of the second-order exchange-correlation
energy with respect to $\tau_{\kappa \lambda}$ (third term).


\subsection{Sum rule}

As in the cases of the Born effective charges and of the dynamical
matrix~\cite{prb1_910}, the coefficients
$\partial \chi_{ij}^{(1)} / \partial \tau_{\kappa \alpha}$
must vanish when they are summed over all atoms in the unit cell.
\begin{equation} \label{eq_sumrule}
\sum_{\kappa}
\frac{\partial \chi_{ij}^{(1)}}{\partial \tau_{\kappa \alpha}}
= 0
\end{equation}
Physically, this sum rule guarantees the fact that the macroscopic
dielectric susceptibility remains invariant under a rigid translation of the
crystal. In practical calculations, it is not always satisfied although the
violation is generally less severe than in case of $\widetilde{C}$ or
$Z^{\ast}$. Even in calculations that present a low degree of convergence,
the deviations from this law can be quite weak. They can be corrected using
similar techniques as in case of the Born effective
charges~\cite{xgcl}. For example, we can define
the mean excess of
$\partial \chi_{ij}^{(1)} / \partial \tau_{\kappa \alpha}$
per atom
\begin{equation}
\overline{
\frac{\partial \chi_{ij}^{(1)}}{\partial \tau_{\alpha}}
}
=
\frac{1}{N_{at}}
\sum_{\kappa}
\frac{\partial \chi_{ij}^{(1)}}{\partial \tau_{\kappa \alpha}}
\end{equation}
and redistribute it equally between the atoms
\begin{equation}
\frac{\partial \chi_{ij}^{(1)}}{\partial \tau_{\kappa \alpha}}
\rightarrow
\frac{\partial \chi_{ij}^{(1)}}{\partial \tau_{\kappa \alpha}}
-
\overline{
\frac{\partial \chi_{ij}^{(1)}}{\partial \tau_{\alpha}}
}.
\end{equation}


\subsection{Electrooptic tensor} \label{sec_eo}

The optical properties of a compound usually depend on external
parameters such as the temperature, electric fields or mechanical
constraints (stress, strain).
In the present section we consider the variations of the refractive
index induced by a static or low-frequency electric field
$\mathcal{E}_{\gamma}$. At the linear order, these variations are
described by the linear EO coefficients
(Pockels effect)
\begin{equation} \label{eq_eo_def}
\Delta \left ( \varepsilon^{-1} \right )_{ij}
=
\sum_{\gamma = 1}^{3}
r_{ij \gamma} \mathcal{E}_{\gamma}
\end{equation}
where $(\varepsilon^{-1})_{ij}$ is the inverse of the electronic
dielectric tensor and $r_{ij \gamma}$ the EO tensor.

Within the Born and Oppenheimer approximation, the EO tensor can be
expressed as the sum of three contributions:
a bare electronic part
$r_{ij \gamma}^{el}$,
an ionic contribution
$r_{ij \gamma}^{ion}$
and a piezoelectric contribution
$r_{ij \gamma}^{piezo}$.

The electronic part is due to an interaction of $\mathcal{E}_{\gamma}$ with the
valence electrons when considering the ions artificially as clamped at
their equilibrium positions. It can be computed from the non-linear optical
coefficients. As can be be seen from Eq. (\ref{eq_nlo_pol}),
$\chi_{ijl}^{(2)}$ defines the second-order change of the
induced polarization with respect to $\mathcal{E}_{\gamma}$.
Taking the derivative
of Eq. (\ref{eq_nlo_pol}), we also see that
$\chi_{ijl}^{(2)}$
defines the first-order change of the linear dielectric susceptibility, which
is equal to
$
\frac{1}{4 \pi}
\Delta \varepsilon_{ij}.
$
Since the EO tensor depends on $\Delta (\varepsilon^{-1})_{ij}$
rather than $\Delta \varepsilon_{ij}$, we have to transform
$\Delta \varepsilon_{ij}$ by the inverse of the zero field electronic
dielectric tensor~\cite{sanchez}
\begin{equation} \label{eq_transfo}
\Delta (\varepsilon^{-1})_{ij} =
-\sum_{m,n=1}^{3}
\varepsilon^{-1}_{im}
\Delta (\varepsilon)_{mn}
\varepsilon^{-1}_{nj}.
\end{equation}
Using Eq. (\ref{eq_transfo}) we obtain the following expression for the
electronic EO tensor
\begin{equation} \label{eq_eo_el}
r_{i j \gamma}^{el} =
- 8 \pi \sum_{l,l'=1}^{3}
\left .
(\varepsilon^{-1})_{il}
\chi^{(2)}_{l l' k}
(\varepsilon^{-1})_{l' j}
\right |_{k=\gamma}.
\end{equation}
Eq. (\ref{eq_eo_el}) takes a simpler form when expressed in the
principal axes of the crystal under investigation~\cite{principal_axes}
\begin{equation} \label{eq_eo_el1}
r_{i j \gamma}^{el} =
\left .
\frac{-8 \pi}{n_i^2 n_j^2} \chi^{(2)}_{ij k}
\right |_{k=\gamma}
\end{equation}
where the $n_i$ coefficients are the principal refractive indices.

The origin of the ionic contribution to the EO tensor is the relaxation of the
atomic positions due to the applied electric field $\mathcal{E}_{\gamma}$ and the variations of
$\varepsilon_{ij}$ induced by these displacements.
It can be computed from the Born effective charges
$Z^{\ast}_{\kappa, \alpha \beta}$
and the
$\frac{\partial \chi_{ij}}{\partial \tau_{\kappa \alpha}}$
coefficients introduced in
Sec. \ref{sec_raman}.
As shown in appendix \ref{sec_app}
[see also Refs. \onlinecite{prb1_3494,pr160_519}],
the ionic EO tensor can be
computed as a sum over the transverse optic phonon modes at $\bm{q} = 0$
\begin{equation} \label{eq_eo_ion}
r_{ij \gamma}^{ion} =
- \frac{4 \pi}{\sqrt{\Omega} n_i^2 n_j^2}
\sum_m
\frac{\alpha_{ij}^m p_{m,\gamma}}{\omega_m^2}
\end{equation}
where $\bm{\alpha}^m$ is the Raman susceptibility of mode $m$
[Eq. (\ref{eq_ramansus})] and $p_{m,\beta}$
the mode polarity
\begin{equation} \label{eq_eo_pol}
p_{m,\gamma} =
\sum_{\kappa,\beta}
Z^{\ast}_{\kappa, \gamma \beta}
u_m(\kappa \beta).
\end{equation}
that is directly linked to the modes oscillator strength
\begin{equation}
S_{m,\alpha \beta}=p_{m, \alpha} \cdot p_{m, \beta}.
\end{equation}
For simplicity, we have expressed Eq. (\ref{eq_eo_ion})
in the principal axes while a more general expression can be derived from
Eq. (\ref{eq_transfo}).

Finally, the piezoelectric contribution is due to a relaxation of the unit
cell shape due to the converse piezoelectric
effect~\cite{veithen_unpublished,asss3_264}.
As it is discussed in appendix \ref{sec_app},
it can be computed from the elasto-optic coefficients
$p_{ij \mu \nu}$ and the piezoelectric strain coefficients
$d_{\gamma \mu \nu}$
\begin{equation} \label{eq_piezo}
r_{ij \gamma}^{piezo} =
\sum_{\mu, \nu = 1}^{3} p_{ij \mu \nu} d_{\gamma \mu \nu}.
\end{equation}

In the discussion of the EO effect, we have to specify
whether we are dealing with strain-free (clamped) or stress-free
(unclamped) mechanical boundary conditions. The clamped EO tensor
$r_{ij \gamma}^{\eta}$
takes into account the electronic and ionic contributions
but neglects any modification of
the unit cell shape due to the converse
piezoelectric effect~\cite{veithen_unpublished,asss3_264}
\begin{equation}
r_{ij \gamma}^{\eta} = r_{ij \gamma}^{el} + r_{ij \gamma}^{ion}.
\end{equation}
Experimentally, it
can be measured for frequencies of $\mathcal{E}_{\gamma}$ high enough
to eliminate the relaxations of the crystal lattice but low enough
to avoid excitations of optical phonon modes
(usually above $\sim$ 100 MHz).
To compute the unclamped EO tensor $r_{ij \gamma}^{\sigma}$,
we have to add the piezoelectric
contribution to $r_{ij \gamma}^{\eta}$
\begin{equation}
r_{ij \gamma}^{\sigma} = r_{ij \gamma}^{\eta} + r_{ij \gamma}^{piezo}.
\end{equation}
Experimentally, $r_{ij \gamma}^{\sigma}$ can be measured for
frequencies of $\mathcal{E}_{\gamma}$ below the (geometry dependent)
mechanical body resonances of the sample~\cite{asss3_264}
(usually below $\sim$ 1 MHz).

\section{Results} \label{sec_res}

\subsection{Technical details}

Our calculations have been performed within the local density approximation
(LDA) to the density functional theory~\cite{hk,ks} (DFT).
We used the {\sc abinit}~\cite{abinit} package,
a planewave, pseudopotential DFT code~\cite{bbbcg}
in which we have implemented
the formalism presented above.
For reasons that will become obvious below, we
chose the PEAD formulation Eq. (\ref{eq_pead})
to perform the differentiation
with respect to \textbf{k}. For the exchange-correlation energy
$E_{xc}$ we relied on the parametrization of Perdew and Wang~\cite{prb45_13244}
as well as the
parametrization of Goedecker, Teter and Hutter~\cite{prb54_1703}.
These expressions have the advantage to avoid any discontinuities in the
functional derivative of $E_{xc}$.

In case of the semiconductors Si, AlAs and AlP,
we used at $16 \times 16 \times 16$
grid of special k-points, a plane-wave kinetic energy cutoff of 10
hartree and Troullier-Martins~\cite{prb43_1993} norm-conserving
pseudopotentials. These
calculations have been performed at the theoretical lattice constant.
To perform the finite difference calculations of the Raman polarizabilities,
changes of the electronic dielectric tensor were computed
for atoms displaced by $\pm 1 \%$ of the unit cell parameter along
the cartesian directions.

In case of rhombohedral BaTiO$_3$, we used a $10 \times 10 \times 10$
grid of special k-points, a plane-wave kinetic energy cutoff of 45
hartree and extended norm-conserving pseudopotentials~\cite{prb48_5031}.
Since the ferroelectric instability is quite sensitive to the
volume of the unit cell
and tends to disappear due to the volume underestimation of the
LDA~\cite{nature358_137}, we chose to work at the experimental lattice
constants. At the opposite to the lattice parameters, the atomic
positions have been relaxed : the residual forces on the atoms
were smaller than $5 \cdot 10^{-5}$ hartree/bohr.

It was shown by Gonze, Ghosez and Godby~\cite{prl74_4035} that an accurate
functional for the exchange-correlation energy in extended systems
should depend on both the density and the polarization. The LDA
used here neglects this polarization dependence and may consequently
indroduce significant relative errors when studying the response of a
solid to an electric field. 
In case of the second-order derivatives,
the LDA usually yields an overestimate of the dielectric tensor (as large
as 20 \% in BaTiO$_{3}$)~\cite{epl33_713}.
At the opposite, no clear trends have been reported yet concerning
non-linear optical properties such as 
$\chi_{ijl}^{(2)}$.~\cite{prl66_41,prb53_15638}


In LDA calculations, it is common practice to apply a {\it scissors
correction}~\cite{prl63_1719} to compensate the lack of polarization
dependence of the exchange-correlation functional. In case of non-linear
optical properties, such a correction can be applied at different levels.
On the one hand, we can compute the non-linear optical susceptibilities
(Eq. (\ref{eq_e3_elfd})) using a scissors operator
for the first-order wavefunctions~\cite{xgcl}.
On the other hand, in the computation of the EO coefficients, we can use
a scissors corrected refractive index in Eqs (\ref{eq_eo_el1}) and
(\ref{eq_eo_ion}). The influence of these corrections will be discussed
below.


\subsection{Non-linear optical susceptibilities  and Raman polarizabilities
of semiconductors} \label{sec_res_sc}

In order to illustrate the computation of third-order energy derivatives, we
performed a series of calculations on various cubic
semiconductors. In these compounds, the non-linear optical
susceptibility tensor $d_{ijk}$ and the Raman susceptibility tensor
$\alpha_{ij}$ only have one independent element d$_{123}$ and $\alpha_{12}$.
Also, instead of $\alpha_{12}$
it is customary to report the Raman polarizability~\cite{cardona} defined as
\begin{equation} \label{eq_ramanpol}
a = \sqrt{\mu \Omega} \alpha_{12}
\end{equation}
where $\mu$ is the reduced mass of the two atoms in the unit cell.

\begin{figure}[htb]
\begin{center}
\includegraphics[width=6cm]{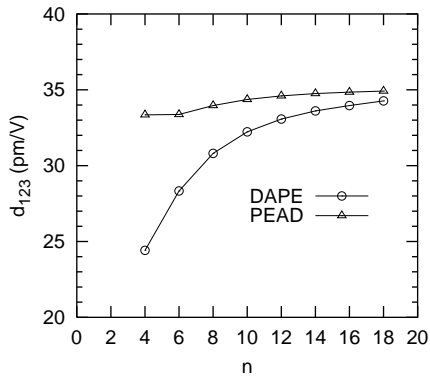}
\end{center}
\caption{\label{fig_convergence} Non-linear optical susceptibility
d$_{123}$ (pm/V) of AlAs for various grids of
$n \times n \times n$ special {\bf k}-points.}
\end{figure}

The formalism of Sec. \ref{sec_form} involves an integration over the BZ
and a differentiation with respect to \textbf{k}. In practical calculations,
these operations must be performed on a discrete mesh of special
\textbf{k}-points. As we explained in Sec. \ref{sec_form}, the discretization
can either be performed before (PEAD) or after (DAPE) the perturbation
expansion  of the energy functional Eq. (\ref{eq_energy}).
Up to know, the applications of the present formalism to real
materials~\cite{prb66_100301,prb53_15638}
made use of the DAPE formula of the third-order energy.
The only application of the PEAD formula has been reported by Nunes and
Gonze~\cite{prb63_155107} on a one-dimensional model system. These authors
observed that the PEAD formula converges better with respect to the
\textbf{k}-point sampling than the DAPE formula. In order to compare the
performance of these two approaches on a realistic case, we applied both of
them to compute the non-linear optical susceptibility d$_{123}$ of AlAs.
We
performed a series of calculations on a $n \times n \times n$ grid of
special \textbf{k}-points. As can be seen on Figure \ref{fig_convergence}
the PEAD formula converges much faster than the DAPE formula.
Therefore, the PEAD formulation has been applied to obtain the results
presented below. It is the one that is actually available in the {\sc
abinit} code.

In Table \ref{tab_scnl}, we report the non-linear
optical susceptibilities of the cubic semiconductors AlAs and AlP.
Our results are in close agreement with the values obtained by Dal Corso
and co-workers~\cite{prb53_15638} who applied the $2n + 1$ theorem within
the DAPE formalism, the results of Levine and
co-workers~\cite{prl66_41} who used a ``sum over excited
states'' technique and the values obtained by Souza and
co-workers~\cite{prl89_117602}
who followed a finite electric field approach.
The values in the lower part of Table \ref{tab_scnl}
have been obtained using a scissors correction. Our methodology
provides a correction similar to what is reported by Levine and
Allan~\cite{prb44_12781}.

The scissors correction decreases the value of the non-linear
optical susceptibilities
in agreement with the discussion of Ref. \onlinecite{prb54_8540}.
To the authors knowledge, no
experimental data are available for AlAs and AlP. For other cubic
semiconductors, it is however not clear that the use of a scissors
correction improves the agreement with the experiment~\cite{prl66_41}
and will even have a negative effect when the LDA underestimates
the experimental value. In addition, 
it is not straightforward to isolate the error of the LDA on the
non-linear response functions from other sources of errors.
Other factors have a similar strong influend on $\chi_{ijl}^{(2)}$
as the scissors correction. For example, the values of 
the non-linear optical susceptibilities strongly depend 
on the pseudopotential~\cite{prb53_15638}
or on the error on the unit cell volume~\cite{prl66_41,prb54_8540} 
that is
usually underestimated in LDA calculations.

\begin{table}[htbp]
\caption{\label{tab_scnl} Non-linear optical susceptibilities $d_{123}$
(pm/V) of some semiconductors. The values in the lower part of
the table have been obtained using a scissors (SCI) correction.}
\begin{ruledtabular}
\begin{tabular}{llcc}
Method  & &  AlAs  &  AlP  \\
\hline
$2n+1$ theorem (present)& &
  35 & 21  \\
$2n+1$ theorem~\footnotemark[1] & &
  32 & 19 \\
Finite fields~\footnotemark[2] & &
  32 & 19  \\
Sum over states~\footnotemark[3] & &
  34 & 21 \\
\hline
$2n+1$ theorem + SCI (present) & &
  21 & 13 \\
Sum over states + SCI~\footnotemark[3] & &
  21 & 13 \\
\end{tabular}
\end{ruledtabular}
\footnotetext[1]{Ref. \onlinecite{prb53_15638}}
\footnotetext[2]{Ref. \onlinecite{prl89_117602}}
\footnotetext[3]{Ref. \onlinecite{prb44_12781}}
\end{table}


We also computed the Raman polarizabilities of
the transverse (TO) and longitudinal optical (LO)
phonons of various semiconductors.
In addition, we performed finite difference calculations of the dielectric
tensor with respect to atomic displacements. Our results are summarized in
Table \ref{tab_scraman} where we also report the results of Deinzer and
Strauch~\cite{prb66_100301} (DS), Baroni and Resta~\cite{prb33_5969} (BR)
as well as the experimental result of
Wagner and Cardona~\cite{ssc48_301} for Si.
The agreement between our results and those obtained in previous works is
quite good. In addition, the results we obtained from the $2n + 1$ theorem
closely agree with the finite difference calculations
giving us some indication of the numerical accuracy of the implementation.

The Raman polarizabilities of the TO and LO modes are different.
As it is discussed in Sec. \ref{sec_raman}, this difference is attributed
to the macroscopic electric field associated to a longitudinal polar
lattice vibration. On the one hand, this field modifies the dynamical
matrix at $\textbf{q} \rightarrow 0$.
The eventual related modification of the
eigenvectors of the LO modes may imply a first
change of the Raman susceptibility.
On the other hand, the macroscopic electric
field itself may induce an additional change of
$\alpha$ related to the non-linear
optical coefficients $\chi_{ijl}^{(2)}$.
In the cubic semiconductors, the
eigenvectors of the TO and LO modes are identical.
The difference between
the polarizabilities of the TO and LO modes comes
therefore exclusively from the
second term of Eq. (\ref{eq_dchidtau_long}).

\begin{table*}[htbp]
\caption{\label{tab_scraman} Raman polarizabilities of the transverse (TO)
and longitudinal (LO) optical modes (\AA$^2$) of some
semiconductors.}
\begin{ruledtabular}
\begin{tabular}{ccccccc}
                   & &  Si   &  AlAs (TO) & AlAs (LO) & AlP (TO) & AlP (LO) \\

\hline
$\bm{2n+1}$ {\bf Theorem} \\
Present           & & 20.02 & 8.48       & 12.48     & 4.30     & 7.46     \\
DS~\footnotemark[1]
                   & & 23.56 & 7.39       &           & 5.13     &          \\
\hline
{\bf Finite differences}\\
Present           & & 20.17 & 8.59       &           & 4.25     &          \\
DS~\footnotemark[1]
                   & & 20.44 & 5.64       &           & 4.44     &          \\
BR~\footnotemark[2]
                   & & 26.16 \\
\hline
{\bf Experiment}
                   & & 23 $\pm$ 4~\footnotemark[3]
                     &  
                     &  
\end{tabular}
\end{ruledtabular}
\footnotetext[1]{Ref. \onlinecite{prb66_100301}}
\footnotetext[2]{Ref. \onlinecite{prb33_5969}}
\footnotetext[3]{Ref. \onlinecite{ssc48_301}}
\footnotetext[4]{Ref. \onlinecite{pr188_1209}}
\footnotetext[5]{Ref. \onlinecite{pma45_239}}
\end{table*}


\subsection{EO tensor in ferroelectric oxides} \label{sec_eo_bto}

In the rombohedral phase of BaTiO$_3$, the EO tensor has four
independent elements: r$_{13}$, r$_{33}$, r$_{22}$ and r$_{51}$.
At the opposite to the dielectric tensor, the EO coefficients can either
be positive or negative. The sign of these coefficients is often
difficult to measure experimentally. Moreover, it depends on the choice of
the cartesian axes. Experimentally, these axes are chosen according to the
{\it Standards on Piezoelectric crystals}. The $z$-axis is along the
direction of the spontaneous polarization and the $y$-axis lies in a mirror
plane. The $z$ and $y$ axis are both piezoelectric. Their positive ends
are chosen in the direction that becomes negative under compression.
The orientation of these axes can easily be found from pure geometrical
arguments. Unfortunately, these arguments do not allow to determine the
direction of the $y$-axis. Therefore, we applied the methodology of
Ref.~[\onlinecite{jpcs61_147}] to compute the piezoelectric tensor from finite
differences of the Berry phase polarization. Our results are reported in
the frame where the piezoelectric coefficient e$_{22}$ and e$_{33}$ are
positive.

These coefficients, as well as their decomposition on the individual phonon
modes and their electronic part, are reported in table \ref{tab_eorhombo}.
All EO coefficients are positive. As it is the case for the tetragonal
phase~\cite{veithen_unpublished}, the modes that have the strongest overlap
with the soft mode of the paraelectric phase dominate the amplitude to the
EO coefficients. Moreover, the electronic contribution is found to be quite
small.

\begin{table}[htbp]
\caption{\label{tab_eorhombo} Decomposition of the clamped EO tensor
(pm/V) in the rhombohedral phase of BaTiO$_3$. Reported are the
contributions
of individual zone-center phonon modes and the electronic contribution.
The phonon frequencies are
reported in cm$^{-1}$.}
\begin{ruledtabular}
\begin{tabular}{crrccrr}
\multicolumn{3}{c}{A$_1$-modes} & & \multicolumn{3}{c}{E-modes} \\
$\omega$ & r$_{13}^{\eta}$ & r$_{33}^{\eta}$  & & 
$\omega$ & r$_{22}^{\eta}$ & r$_{51}^{\eta}$ \\
\hline
168      &  0.65 &  2.16 & & 163 &  0.79 &  5.15  \\
253      & 13.82 & 27.32 & & 202 &  5.40 & 19.16  \\
509      &  1.31 &  2.05 & & 293 &  0.01 & -0.02  \\
         &       &       & & 469 &  0.24 &  0.65  \\
\hline
Elect.   &  1.15 &  2.95 & &     &  0.12 &  1.24  \\
\hline
Tot.     & 16.93 & 34.48 & &     &  6.56 & 26.18  \\
\end{tabular}
\end{ruledtabular}
\end{table}

%

As we discussed in the previous sections, 
linear and non-linear optical susceptibilities are
sometimes relatively inaccurate within the LDA. In this context,
it is interesting to investigate the error due to the use of the
LDA optical dielectric constants in the transformation
Eq. (\ref{eq_transfo}). Unfortunately, we could not find any experimantal
data on the EO coefficients in the rhombohedral phase of
BaTiO$_3$. In Ref. \onlinecite{veithen_unpublished},
we studied the EO coefficients of ferroelectric LiNbO$_3$
and tetragonal BaTiO$_3$ and PbTiO$_3$ and found an overall
good agreement between theory and experiment.
In Table \ref{tab_sci}, we report the EO coefficients of
these compounds as well as the values obtained using a scissors
corrected optical dielectric constant. 
No scissors correction
has been applied for the non-linear optical susceptibilities of 
these compounds that are required to compute the electronic
contributions.

The effect of this correction is more important for the perovskite
compounds than for LiNbO$_{3}$, for which the LDA bandgap and optical
dielectric constants are in reasonable agreement with the
experiment~\cite{prb65_214302} .
For BaTiO$_{3}$, we tested the optical dielectric
tensor obtained from the scissors correction that modifies the LDA bandgap
to its experimental value~\cite{xgcl}: we obtain r$_{13}^{\eta}$ = 
12.68 pm/V and
r$_{33}^{\eta}$ =
30.84 pm/V in closer agreement with experimental data. However, such an
improvement is not a general rule.
In PbTiO$_3$, a scissors shift that correct the LDA bandgap
fails to correct the LDA optical dielectric constant
(we obtain $\varepsilon_{11}$ = 5.81 and $\varepsilon_{33}$ = 5.51
while the experimental values are 6.63 and 6.64~\cite{prb48_10160})
and yields r$_{13}^{\eta}$ = 14.24 pm/V and r$_{33}^{\eta}$ = 8.94 pm/V.
Using the experimental dielectric constants, we obtain
r$_{13}^{\eta}$ = 10.92 pm/V and r$_{33}^{\eta}$ = 6.16 pm/V in better
agreement with the experiment.

\begin{table}[htbp]
\caption{\label{tab_sci} Effect of a scissors correction on the EO
coefficients of LiNbO$_3$ (clamped and unclamped cases), and the
tetragonal phases of PbTiO$_3$ and BaTiO$_3$ (clamped cases only).
The dielectric tensor required to perform the transformation
Eq. (\ref{eq_transfo}) has been computed whitin the LDA 
($\varepsilon_{LDA}$) and using the LDA with a scissors correction
($\varepsilon_{SCI}$). No scissors correction has been used
to compute the non-linear optical susceptibilities that determine
the electronic contribution to the EO coefficients.
In case of PbTiO$_3$, we also use the
experimental dielectric tensor ($\varepsilon_{exp}$) to compute
the EO coefficients. The values are compared to the experimental
results.
}
\begin{ruledtabular}
\begin{tabular}{llrrrr}
           &                     & r$_{13}$ & r$_{33}$ & r$_{22}$ & r$_{51}$ \\
\hline
LiNbO$_3$  & $\varepsilon_{LDA}$ &  9.67    & 26.93    &  4.55    & 14.93 \\
(clamped)  & $\varepsilon_{SCI}$ & 10.37    & 28.89    &  4.88    & 16.02 \\
           & Exp.~\cite{rauber}  &  8.6     & 30.8     &  3.4     & 28    \\
\hline
LiNbO$_3$  & $\varepsilon_{LDA}$ & 10.47    & 27.08    &  7.53    & 28.61 \\
(unclamped)& $\varepsilon_{SCI}$ & 11.23    & 29.06    &  8.08    & 30.69 \\
           & Exp.~\cite{rauber}  & 10       & 32.2     &  6.8     & 32.6  \\
           & Exp.~\cite{jap84_2251}&        &          &  9.89    &       \\
\hline
PbTiO$_3$  & $\varepsilon_{LDA}$ &  8.98    &  5.88    &          & 30.53 \\
(clamped)  & $\varepsilon_{SCI}$ & 14.24    &  8.94    &          & 47.39 \\
           & $\varepsilon_{exp}$ & 10.92    &  6.16    &          & 34.45 \\
           & Exp.~\cite{handbook_laser}&13.8&  5.9     &          &       \\
\hline
BaTiO$_3$  & $\varepsilon_{LDA}$ &  8.91    & 22.27    &          &       \\
(clamped)  & $\varepsilon_{SCI}$ & 12.68    & 30.84    &          &       \\
           & Exp.~\cite{prb50_5941}&10.2    & 40.6     &          &       \\
           & Exp.~\cite{asss3_264} & 8      & 28       &          &       \\
\end{tabular}
\end{ruledtabular}
\end{table}


\section{Conclusions and perspectives}

In this paper, we presented the general framework for the
computation of third-order
energy derivatives within DFT.
Our formalism makes use of the $2n+ 1$ theorem and the
modern theory of polarization. Focusing on derivatives that are
characterized by a zero wavevector and that involve either three electric
fields or two electric fields and one atomic displacement, we described the
computation of non-linear optical susceptibilities, of Raman scattering
efficiencies of TO and LO phonons and of the EO tensor.

The computation of the Berry phase polarization involves a derivative
of the wavefunctions with respect to their wavevector.
In practice, this differentiation
is computed  on a grid of special
\textbf{k}-points. The perturbation expansion can either be performed before
(DAPE) or after (PEAD) the discretization, leading to two mathematically
distinct expression of the third-order energies. We used both of them
to compute the non-linear optical susceptibility of AlAs and we have shown
that the PEAD formulation converges faster with respect to the
\textbf{k}-point sampling.

We have computed the
non-linear optical susceptibilities and Raman polarizabilities of some
cubic semiconductors as well as the EO tensor in the
rhombohedral phase of BaTiO$_3$.

Finally, we have studied the effect of a scissors correction on
the EO coefficients and the non-linear optical susceptibilities.
At the opposite to the dielectric tensor, we did not find a systematic
improvement of the results by using this correction.

We can figure out several applications of the methodology presented in this
work.
Combined with the calculation of phonon frequencies and infrared
intensities,
the computation of Raman efficiencies can be a useful
complementary tool for
the interpretation of experimental spectra. Furthermore, the computation
of the EO tensor from first-principles can guide the tuning of the EO
properties and help designing new
efficient EO materials.
This could reveal particularly helpful since accurate optical measurements
require high quality single crystals not always directly accessible.

\section{Acknowledgments}
The authors are grateful to M. D. Fontana, P. Bourson, B. Kirtman and
B. Champagne for helpful discussions. M. V. and X. G. acknowledge
financial support from the FNRS Belgium. This work was supported by
the Volkswagen-Stiftung within the project ``Nano-sized ferroelectric
Hybrids" (I/77 737), the Region Wallonne (Nomade, project 115012),
the Communaut\'e Francaise de Belgique - Actions de Recherche Concert\'ees,
the PAI/UIAP Phase 5 "Quantum size effects in nanostructured materials",
FNRS-Belgium through grants 9.4539.00 and 2.4562.03,
the European Union through the Research and Training Network "EXCITING"
(HPRM-CT-2002-00317).


\appendix

\section{Expressions of the clamped and unclamped EO tensors} \label{sec_app}

\subsection{Macroscopic approach} \label{sec_app_mac}

As it is discussed in Sec. \ref{sec_eo}, the optical properties
of a compound are modified by an electric field
$\mathcal{E}_{\gamma}$ or a mechanical constraint (a stress
$\sigma_{\mu \nu}$ or a homogeneous strain $\eta_{\mu \nu}$).
At the linear order, the variations of
$\varepsilon_{ij}^{-1}$ can be described using either the variables
$(\mathcal{E}_{\gamma},\eta_{\mu \nu})$ or
$(\mathcal{E}_{\gamma},\sigma_{\mu \nu})$~\cite{jap78_2651,nye}
\begin{subequations}
\begin{eqnarray}
\Delta(\varepsilon^{-1})_{ij} & = &
\sum_{\gamma = 1}^{3} r_{ij \gamma}^{\eta} \mathcal{E}_{\gamma} +
\sum_{\mu,\nu = 1}^{3} p_{ij \mu \nu}
\eta_{\mu \nu} \label{eq_reta} \\
\Delta(\varepsilon^{-1})_{ij} & = &
\sum_{\gamma = 1}^{3} r_{ij \gamma}^{\sigma} \mathcal{E}_{\gamma} +
\sum_{\mu,\nu = 1}^{3} \pi_{ij \mu \nu}
\sigma_{\mu \nu} \label{eq_rsigma}
\end{eqnarray}
\end{subequations}
where $r_{ij \gamma}^{\eta}$ and $r_{ij \gamma}^{\sigma}$ are resp.
the clamped (strain free) and unclamped (stress free) EO coefficients,
$p_{ij \mu \nu}$ are the elasto-optic (strain-optic) coefficients
and $\pi_{ij \mu \nu}$ are the piezo-optical (stress-optical)
coefficients.
In order to relate Eqs. (\ref{eq_reta}) and (\ref{eq_rsigma}),
we can express the strain as beeing induced by the stress or by
the electric field (converse piezoelectric effect)
\begin{equation} \label{eq_strain}
\eta_{\mu \nu} =
\sum_{\mu',\nu' = 1}^{3} S_{\mu \nu \mu' \nu'} \sigma_{\mu' \nu'}
+
\sum_{\gamma = 1}^{3} d_{\gamma \mu \nu} \mathcal{E}_{\gamma}
\end{equation}
where $S_{\mu \nu \mu' \nu'}$ are the elastic compliances
and $d_{\gamma \mu \nu}$ the piezoelectric strain coefficients.

If we assume for example that the unit cell is free to relax
within the electric field (stress-free mechanical boundary conditions)
we can either use Eq. (\ref{eq_rsigma}) (in which case the second
term of the right-hand-side is zero) or Eq. (\ref{eq_reta})
to compute $\Delta(\varepsilon^{-1})_{ij}$. In the latter case,
the strain induced by the electric field can be obtained
from the second term of the right-hand-side of Eq. (\ref{eq_strain})
\begin{eqnarray}
\Delta(\varepsilon^{-1})_{ij} & = &
\sum_{\gamma = 1}^{3} r_{ij \gamma}^{\sigma} \mathcal{E}_{\gamma} \nonumber \\
& = &
\sum_{\gamma = 1}^{3} r_{ij \gamma}^{\eta} \mathcal{E}_{\gamma} +
\sum_{\mu,\nu = 1}^{3} \sum_{\gamma = 1}^{3}
p_{ij \mu \nu} d_{\gamma \mu \nu} \mathcal{E}_{\gamma}.
\end{eqnarray}
Using this identity, we obtain the following relation between
the unclamped and the clamped EO coefficients
\begin{equation}
r_{ij \gamma}^{\sigma} = r_{ij \gamma}^{\eta} +
\sum_{\mu,\nu = 1}^{3} p_{ij \mu \nu} d_{\gamma \mu \nu}.
\end{equation}

\subsection{Microscopic approach} \label{sec_ent}

In order to derive the expressions of the
clamped and unclamped EO tensor of Sec. \ref{sec_eo},
we use a
Taylor expansion of the electric enthalpy~\cite{landlif} $F$.
Similar developments have already been applied to determine the lattice
contribution of the static dielectric tensor and of the piezoelectric
tensor~\cite{umesh,dc_phd}. They are based on an expansion of $F$ up
to the second order in the atomic coordinates
$R_{\kappa \alpha}$, the homogeneous
strain $\eta_{\mu \nu}$
and the macroscopic electric field $\mathcal{E}_{\gamma}$.
In this section, we extend
these developments to the third order.

The electric enthalpy of a solid in an electric
field is obtained by the minimization
\begin{equation} \label{eq_ion_min}
F(\bm{\mathcal{E}}) = \min_{\textbf{R},\eta}
F \left ( \textbf{R}, \eta, \bm{\mathcal{E}} \right ).
\end{equation}
We denote $\textbf{R}(\bm{\mathcal{E}})$,
$\eta(\bm{\mathcal{E}})$ the atomic positions and the strain
that minimize $F$ at constant $\bm{\mathcal{E}}$ and
$\textbf{R}_0$, $\eta_0$ (= 0) their values
at $\bm{\mathcal{E}}=0$. For small fields, we
can expand the function
$F \left ( \textbf{R}, \eta, \bm{\mathcal{E}} \right )$
in powers of $\bm{\mathcal{E}}$ around $\bm{\mathcal{E}}=0$:
\begin{equation} \label{eq_ion_expansion}
F \left ( \textbf{R}, \eta, \mathcal{E} \right ) =
F \left ( \textbf{R}, \eta, 0 \right )
- \Omega \sum_{i=1}^{3}
P_i \left ( \textbf{R},\eta \right ) \mathcal{E}_i
- \frac{\Omega}{8 \pi}
\sum_{i,j=1}^{3}
\varepsilon_{ij} \left ( \textbf{R},\eta \right )
\mathcal{E}_i \mathcal{E}_j
- \frac{\Omega}{3}
\sum_{i,j,k=1}^{3}
\chi_{ijk}^{(2)} \left ( \textbf{R},\eta \right )
\mathcal{E}_i \mathcal{E}_j \mathcal{E}_k
+ \cdots
\end{equation}
where
$\Omega$ is the volume of the primitive unit cell in real space and
$P \left ( \textbf{R}, \eta \right )$,
$\varepsilon_{ij} \left ( \textbf{R},\eta \right )$ and
$\chi_{ijk}^{(2)} \left ( \textbf{R},\eta \right )$ are the
macroscopic polarization, electronic dielectric tensor and
non-linear optical coefficients
at zero macroscopic electric field and for a given
configuration (${\bf R}$, $\eta$).
At non-zero field, these quantities are defined as partial
derivatives of $F$ with respect to $\bm{\mathcal{E}}$. For example,
the electric field dependent electronic dielectric tensor can be
computed from the expression
\begin{equation} \label{eq_ion_diel}
\left .
\varepsilon_{ij} \left (
\textbf{R}(\mathcal{E}), \eta(\mathcal{E}), \mathcal{E}
\right ) =
- \frac{4 \pi}{\Omega}
\frac{\partial^2 F}{\partial \mathcal{E}_i \partial \mathcal{E}_j}
\right |_{\textbf{R}(\mathcal{E}),\eta(\mathcal{E}),\mathcal{E}}.
\end{equation}

Let
$
\tau_{\kappa \alpha}=
\textbf{R}_{\kappa \alpha}-\textbf{R}_{0,\kappa \alpha}
$
be the displacement of atom $\kappa$ along direction $\alpha$
and
$\tau_{\kappa \alpha}^{\lambda}$
(resp. $\eta_{\mu \nu}^{\lambda}$)
the first-order modification of the atomic position
(resp. strain) induced by a perturbation $\lambda$
\begin{equation} \label{eq_tau}
\tau_{\kappa \alpha}^{\lambda} =
\left .
\frac{\partial \tau_{\kappa \alpha}}
{\partial \lambda}
\right |_{\lambda = 0},
\hspace{2ex}
\eta_{\mu \nu}^{\lambda} =
\left .
\frac{\partial \eta_{\mu \nu}}
{\partial \lambda}
\right |_{\lambda = 0}.
\end{equation}
In the discussion that follows, we will study the effect of
an electric field perturbation and a strain perturbation on
the electric enthalpy $F$ in order to obtain the formulas
to compute the elasto-optic coefficients as well as the clamped
and the unclamped EO tensors.

\subsubsection{Elasto-optic coefficients ($\bm{\mathcal{E}} = 0$)}

The elasto-optic tensor can be computed from the {\it total} derivative
of the dielectric tensor
with respect to $\eta_{\mu \nu}$ at zero electric field
\begin{equation} \label{eq_depsdeta}
\left .
\frac{d\varepsilon_{ij} \left ( \textbf{R}, \eta, 0 \right )}
{d \eta_{\mu \nu}}
\right |_{\textbf{R}_0, \eta_0}
=
\left .
\frac{\partial \varepsilon_{ij}
\left ( \textbf{R}, \eta \right )}
{\partial \eta_{\mu \nu}}
\right |_{\bm{R}_0, \eta_0}
+
\left .
4 \pi \sum_{\kappa \alpha}
\frac{\partial \chi_{ij}^{(1)} \left ( \textbf{R}, \eta \right )}
{\partial \tau_{\kappa \alpha}}
\right |_{\bm{R}_0, \eta_0}
\tau_{\kappa \alpha}^{\eta_{\mu \nu}}.
\end{equation}
The derivative in the first term of the right hand side
is computed considering the ionic cores
as artificially clamped at their equilibrum positions.
The remaining terms represent the ionic
contribution to the elasto-optic tensor.
They involve derivatives of the linear
dielectric susceptibility $\chi_{ij}^{(1)}$ with respect to the atomic
positions that have to be multiplied by the first-order strain
induced atomic displacements $\tau_{\kappa \alpha}^{\eta_{\mu \nu}}$
[Eq. (\ref{eq_tau})].
To compute these quantities we use the fact that
$F$ is minimum at the
equilibrum for an imposed strain $\eta$. This condition implies
\begin{equation} \label{eq_tau_eq}
\left .
\frac{\partial F \left ( \textbf{R}, \eta \right )}
{\partial \tau_{\kappa \alpha}}
\right |_{\textbf{R}(\eta),\eta}
= 0.
\end{equation}
Since we are interested in first-order atomic displacements we can
write
$
\tau_{\kappa \alpha} (\eta) =
\sum_{\mu, \nu = 1}^{3}
\tau_{\kappa \alpha}^{\eta_{\mu \nu}} \eta_{\mu \nu}
+ \mathcal{O}(\eta^2).
$
Solving the extremum equation (\ref{eq_tau_eq}) to the linear order
in $\eta$, we obtain
\begin{equation} \label{eq_tau_disp}
\sum_{\kappa',\alpha'}
\left .
\frac{\partial^2 F \left ( \textbf{R}, \eta \right )}
{\partial \tau_{\kappa \alpha} \partial \tau_{\kappa' \alpha'}}
\right |_{\textbf{R}_0, \eta_0}
\tau_{\kappa' \alpha'}^{\eta_{\mu \nu}}
=
- \left .
\frac{\partial^2 F \left ( \textbf{R}, \eta \right )}
{\partial \eta_{\mu \nu} \partial \tau_{\kappa \alpha}}
\right |_{\textbf{R}_0, \eta_0}.
\end{equation}
The second derivatives on the left side of Eq. (\ref{eq_tau_disp})
define the matrix of interatomic force constants at zero macroscopic
electric field which enables the computation of the transverse
phonon frequencies
$\omega_m$ and eigendisplacements
$u_m (\kappa \alpha)$.
By decomposing
$\tau_{\kappa \alpha}^{\eta_{\mu \nu}}$
in the basis of the
zone-center phonon-mode eigendisplacements
\begin{equation} \label{eq_ion_decomp}
\tau_{\kappa \alpha}^{\eta_{\mu \nu}} =
\sum_m \tau_m^{\eta_{\mu \nu}} u_m (\kappa \alpha)
\end{equation}
and using Eqs. (\ref{eq_eigen}), (\ref{eq_norm})
we derive the following expression for the first-order strain induced
atomic displacements
\begin{equation} \label{eq_tau_disp1}
\tau_{m}^{\eta_{\mu \nu}} =
\frac{-1}{\omega_m^2}
\left.
\frac{\partial^2 F \left ( \textbf{R}, \eta \right )}
{\partial \eta_{\mu \nu} \partial \tau_m}
\right |_{\textbf{R}_0, \eta_0}
\end{equation}
where
\begin{equation}
\left .
\frac{\partial^2 F \left ( \textbf{R}, \eta \right )}
{\partial \eta_{\mu \nu} \partial \tau_m}
\right |_{\textbf{R}_0, \eta_0}
=
\sum_{\kappa,\alpha}
\left.
\frac{\partial^2 F \left ( \textbf{R}, \eta \right )}
{\partial \eta_{\mu \nu} \partial \tau_{\kappa \alpha}}
\right |_{\textbf{R}_0, \eta_0}
u_m(\kappa \alpha).
\end{equation}
If we introduce Eqs. (\ref{eq_ion_decomp}) and
(\ref{eq_tau_disp1})
into Eq. (\ref{eq_depsdeta}) and use the definition
of the Raman susceptibility Eq. (\ref{eq_ramansus}) and the transformation
Eq. (\ref{eq_transfo}), we finally obtain the formula to compute
the elastooptic tensor
\begin{equation} \label{eq_elastoopt}
p_{ij \mu \nu} =
\frac{-1}{n_i^2 n_j^2}
\left .
\frac{\partial \varepsilon_{ij}
\left ( \textbf{R}, \eta \right )}
{\partial \eta_{\mu \nu}}
\right |_{\bm{R}_0, \eta_0}
+
\frac{4 \pi}{n_i^2 n_j^2 \sqrt{\Omega}}
\sum_{m}
\frac{\alpha_{ij}^m}{\omega_m^2}
\left .
\frac{\partial^2 F \left ( \textbf{R}, \eta \right )}
{\partial \eta_{\mu \nu} \partial \tau_m}
\right |_{\textbf{R}_0, \eta_0}.
\end{equation}
To simplify, we write Eq. (\ref{eq_elastoopt}) in the
principal axes of the crystal under investigation. A more
general expression can be obtained from Eq. (\ref{eq_transfo}).

Eq. (\ref{eq_elastoopt}) is different from the approach used
previously by Detraux and Gonze to study the elasto-optic tensor
in $\alpha$-quartz~\cite{prb63_115118}.
The authors of Ref. \onlinecite{prb63_115118} used finite differences
with respect to strains
to compute the the total derivative of $\varepsilon_{ij}$.
In their approach, the atoms
where relaxed to their equilibrum positions in the strained configurations.
In case of Eq. (\ref{eq_elastoopt}),
the first term of the right-hand-side is computed at
clamped atomic positions while
the effect of the strain-induced atomic relaxations
is taken into account by the second term.

\subsubsection{Clamped EO coefficients ($\eta = 0$)}

The clamped EO tensor can be computed from the {\it total} derivative
of the electric field dependent dielectric tensor
Eq. (\ref{eq_ion_diel})
with respect to $\mathcal{E}$
\begin{equation} \label{eq_clamp_chidc}
\left .
\frac{d\varepsilon_{ij} \left ( \textbf{R}, \eta_0, \bm{\mathcal{E}} \right )}
{d \mathcal{E}_{\gamma}}
\right |_{\bm{R}_0,\mathcal{E}=0}
=
\left .
\frac{\partial \varepsilon_{ij}
\left ( \textbf{R}_0, \eta_0, \bm{\mathcal{E}} \right )}
{\partial \mathcal{E}_{\gamma}}
\right |_{\mathcal{E}=0}
+
\left .
4 \pi \sum_{\kappa \alpha}
\frac{\partial \chi_{ij}^{(1)}
\left ( \textbf{R}, \eta_0 \right )}
{\partial \tau_{\kappa \alpha}}
\right |_{\bm{R}_0}
\tau_{\kappa \alpha}^{\mathcal{E}_{\gamma}}.
\end{equation}
The derivative in the first term is computed considering the ionic cores
as artificially clamped at their equilibrum positions. This term represents
the bare electronic contribution to the EO tensor
that can be computed from the non-linear optical coefficients
\begin{equation} \label{eq_elec_ent}
\left .
\frac{\partial \varepsilon_{ij}
\left ( \textbf{R}_0, \eta_0, \bm{\mathcal{E}} \right )}
{\partial \mathcal{E}_{\gamma}}
\right |_{\mathcal{E}=0}
=
\left .
8 \pi \chi_{ijk}^{(2)} \right |_{k = \gamma}
\end{equation}
related to a third-order partial derivative of $F$
\begin{equation} \label{eq_ion_el}
\left .
\chi_{ijk}^{(2)} =
\chi_{ijk}^{(2)} \left ( \textbf{R}_0,\eta_0 \right ) =
\frac{-1}{2 \Omega}
\frac{\partial^3 F
\left ( \textbf{R}_0, \eta_0, \bm{\mathcal{E}} \right )
}
{\partial \mathcal{E}_i \partial \mathcal{E}_j \partial \mathcal{E}_k}
\right |_{\mathcal{E}=0}.
\end{equation}
The remaining terms in Eq. (\ref{eq_clamp_chidc}) represent the ionic
contribution to the EO tensor.
They involve derivatives of the linear
dielectric susceptibility $\chi_{ij}^{(1)}$ with respect to the atomic
positions that have to be multiplied by the first-order electric field
induced atomic displacements
$\tau_{\kappa \alpha}^{\mathcal{E}_{\gamma}}$
[Eq. (\ref{eq_tau})]. To obtain these quantities, we proceed
the same way as in case of the elasto-optic tensor. Using
the equilibrum condition
\begin{equation} \label{eq_clamp_eq}
\frac{\partial F}{\partial \tau_{\kappa \alpha}} = 0 =
\left .
\frac{\partial F \left ( \textbf{R}, \eta_0, 0 \right )}
{\partial \tau_{\kappa \alpha}}
\right |_{\bm{R}(\mathcal{E})}
-
\left .
\Omega \sum_{i=1}^{3}
\frac{\partial P_i \left ( \textbf{R},\eta_0 \right ) }
{\partial \tau_{\kappa \alpha}}
\right |_{\bm{R}(\mathcal{E})}
\mathcal{E}_i
-
\left .
\frac{\Omega}{8 \pi} \sum_{i,j=1}^{3}
\frac{\partial \varepsilon_{ij} \left ( \textbf{R},\eta_0 \right ) }
{\partial \tau_{\kappa \alpha}}
\right |_{\bm{R}(\mathcal{E})}
\mathcal{E}_i \mathcal{E}_j
+ \cdots
\end{equation}
and expanding $\tau_{\kappa \alpha}$
to the first-order in the electric field, we obtain
\begin{equation} \label{eq_clamp_disp}
\sum_{\kappa', \alpha'}
\left .
\frac{\partial^2 F \left ( \textbf{R},\eta_0, 0 \right )}
{\partial \tau_{\kappa \alpha} \partial \tau_{\kappa' \alpha'}}
\right |_{\textbf{R}_0}
\tau_{\kappa' \alpha'}^{\mathcal{E}_{\gamma}}
=
\left .
\Omega
\frac{\partial P_{\gamma}
\left ( \textbf{R},\eta_0 \right )
}{\partial \tau_{\kappa \alpha}}
\right |_{\textbf{R}_0}.
\end{equation}
This expression is similar to Eq. (\ref{eq_tau_disp}).
The second-order derivatives of $F$ on the left side are
the interatomic force constatnts and
the derivative of the zero field polarization with respect to
$\tau_{\kappa \alpha}$
on the right side is the Born
effective charge tensor
$Z^{\ast}_{\kappa, \gamma \alpha}$ of atom $\kappa$.
Decomposing
$\tau_{\kappa \alpha}^{\mathcal{E}_{\gamma}}$
in the basis of the
zone-center phonon-mode eigendisplacements [Eq. (\ref{eq_ion_decomp})]
and using the orthononormality constraint Eq. (\ref{eq_norm})
we derive the following expression for the first-order electric field induced
atomic displacements
\begin{equation} \label{eq_clamp_disp1}
\tau_{m}^{\mathcal{E}_{\gamma}} =
\frac{1}{\omega_m^2}
\sum_{\kappa,\alpha} Z^{\ast}_{\kappa, \gamma \alpha} u_m(\kappa \alpha).
\end{equation}
If we introduce Eqs. (\ref{eq_elec_ent})
and (\ref{eq_clamp_disp1})
into Eq. (\ref{eq_clamp_chidc}) we finally obtain the formula to compute
the total derivative of the dielectric tensor
\begin{equation} \label{eq_clamp_chi}
\left .
\frac{d\varepsilon_{ij} \left ( \textbf{R}, \mathcal{E} \right )}
{d \mathcal{E}_{\gamma}}
\right |_{\bm{R}_0, \mathcal{E}=0}
=
\left .
8 \pi \chi_{ijk}^{(2)}
\right |_{k = \gamma}
+
4 \pi \sum_m
\frac{1}{\omega_m^2}
\left (
\sum_{\kappa,\alpha}
\frac{\partial \chi_{ij}^{(1)}(\textbf{R})}
{\partial \tau_{\kappa \alpha}}
u_m(\kappa \alpha)
\right )
\left (
\sum_{\kappa',\beta}
Z^{\ast}_{\kappa', \gamma \beta}
u_m(\kappa' \beta)
\right ).
\end{equation}
Using the definition of the Raman susceptibility [Eq. (\ref{eq_ramansus})],
the mode polarity [Eq. (\ref{eq_eo_pol})]
and the transformation [Eq. (\ref{eq_transfo})] we obtain the expression of the
clamped EO tensor
\begin{equation} \label{eq_clamp_rijgamma}
r_{ij \gamma}^{\eta}
=
\left .
\frac{-8 \pi}{n_i^2 n_j^2}
\chi_{ijl}^{(2)}
\right |_{l = \gamma}
-
\frac{4 \pi}{n_i^2 n_j^2 \sqrt{\Omega}}
\sum_m
\frac{\alpha_{ij}^m p_{m,\gamma}}{\omega_m^{2}}
\end{equation}
As in case of the elasto-optic tensor [Eq. (\ref{eq_elastoopt})],
we have written Eq.
(\ref{eq_clamp_rijgamma}) in the principal axes of the crystal
under investigation.

\subsubsection{Unclamped EO tensor ($\sigma = 0$)}

In order to compute the unclamped EO tensor, we have to take into
account both the electric field induced atomic displacments
$\tau_{\kappa \alpha}^{\mathcal{E}_{\gamma}}$ and
the electric field induced strain
$\eta_{\mu \nu}^{\mathcal{E}_{\gamma}}$
when computing the total
derivative of $\varepsilon_{ij}$
\begin{eqnarray} \label{eq_unclamp_chidc}
\left .
\frac{d\varepsilon_{ij} \left ( \textbf{R}, \eta, \bm{\mathcal{E}} \right )}
{d \mathcal{E}_{\gamma}}
\right |_{\bm{R}_0,\eta_0, \mathcal{E}=0}
= & &
\left .
\frac{\partial \varepsilon_{ij}
\left ( \textbf{R}_0, \eta_0, \bm{\mathcal{E}} \right )}
{\partial \mathcal{E}_{\gamma}}
\right |_{\mathcal{E}=0}
+
\left .
4 \pi \sum_{\kappa \alpha}
\frac{\partial \chi_{ij}^{(1)}
\left ( \textbf{R}, \eta_0 \right )}
{\partial \tau_{\kappa \alpha}}
\right |_{\bm{R}_0}
\tau_{\kappa \alpha}^{\mathcal{E}_{\gamma}} \nonumber \\
&& +
4 \pi \sum_{\mu,\nu = 1}^3
\left .
\frac{\partial \chi_{ij}^{(1)}
\left ( \textbf{R}_0, \eta \right )}
{\partial \eta_{\mu \nu}}
\right |_{\eta_0}
\eta_{\mu \nu}^{\mathcal{E}_{\gamma}}.
\end{eqnarray}
The electronic contribution [first term of Eq. (\ref{eq_unclamp_chidc})]
is the same as for the clamped EO tensor. It can be computed from the
non-linear optical coefficients [Eq. (\ref{eq_elec_ent})].
To compute $\tau_{\kappa \alpha}^{\mathcal{E}_{\gamma}}$
and $\eta_{\mu \nu}^{\mathcal{E}_{\gamma}}$, we can use
an equilibrum condition similar to Eq. (\ref{eq_clamp_eq}) where
we require that the first-order derivatives of $F$ with respect
to $\tau_{\kappa \alpha}$ and $\eta_{\mu \nu}$ vanish.
Expanding $\tau_{\kappa \alpha}$ and $\eta_{\mu \nu}$ to the
first-order in the electric field, we obtain the system of coupled
equations [see also Ref. \onlinecite{mrs718_323}]
\begin{subequations}
\begin{eqnarray}
\sum_{\kappa',\alpha'}
\left .
\frac{\partial^2 F \left ( \textbf{R},\eta, 0 \right )}
{\partial \tau_{\kappa \alpha} \partial \tau_{\kappa' \alpha'}}
\right |_{\textbf{R}_0,\eta_0}
\tau_{\kappa' \alpha'}^{\mathcal{E}_{\gamma}}
+
\sum_{\mu, \nu}
\left .
\frac{\partial^2 F \left ( \textbf{R},\eta, 0 \right )}
{\partial \tau_{\kappa \alpha} \partial \eta_{\mu \nu}}
\right |_{\textbf{R}_0,\eta_0}
\eta_{\mu \nu}^{\mathcal{E}_{\gamma}}
& = &
\left .
\Omega
\frac{\partial P_{\gamma}
\left ( \bm{R}, \eta \right )}{\partial \tau_{\kappa \alpha}}
\right |_{\textbf{R}_0,\eta_0} \label{eq_taueta1}  \\
\sum_{\mu', \nu'}
\left .
\frac{\partial^2 F \left ( \textbf{R},\eta, 0 \right )}
{\partial \eta_{\mu \nu} \partial \eta_{\mu' \nu'}}
\right |_{\textbf{R}_0,\eta_0}
\eta_{\mu' \nu'}^{\mathcal{E}_{\gamma}}
+
\sum_{\kappa',\alpha'}
\left .
\frac{\partial^2 F \left ( \textbf{R},\eta, 0 \right )}
{\partial \tau_{\kappa' \alpha'} \partial \eta_{\mu \nu}}
\right |_{\textbf{R}_0,\eta_0}
\tau_{\kappa' \alpha'}^{\mathcal{E}_{\gamma}}
& = &
\left .
\Omega
\frac{\partial P_{\gamma}
\left ( \bm{R}, \eta \right )}{\partial \eta_{\mu \nu}}
\right |_{\textbf{R}_0,\eta_0} \label{eq_taueta2}
\end{eqnarray}
\end{subequations}
Because of the coupling between
$\tau_{\kappa \alpha}^{\mathcal{E}_{\gamma}}$
and
$\eta_{\mu \nu}^{\mathcal{E}_{\gamma}}$, defined
by the mixed second-order derivatives
$
\frac{\partial^2 F}{\partial \tau_{\kappa \alpha} \eta_{\mu \nu}},
$
the second term of the right hand side of Eq. (\ref{eq_unclamp_chidc})
is different from that of Eq. (\ref{eq_clamp_chidc}).
That means that the sum of the first and second term of Eq.
(\ref{eq_unclamp_chidc}) is not identical  to the clamped
EO coefficients $r_{ij \gamma}^{\eta}$.
Moreover, the third term of Eq. (\ref{eq_unclamp_chidc}) is different
from the piezoelectric contribution of Sec. \ref{sec_app_mac}.

In order to obtain the decomposition of $r_{ij \gamma}^{\sigma}$
into electronic, ionic and piezoelectric contributions
defined previously, we can solve Eq. (\ref{eq_taueta1})
for $\tau_{\kappa \alpha}^{\mathcal{E}_{\gamma}}$.
In the basis of the zone-center phonon mode eigendisplacements we
can write
\begin{equation} \label{eq_unclamp_tau}
\tau_n^{\mathcal{E}_{\gamma}} =
\frac{p_{n,\gamma}}{\omega_n^2}
-
\frac{1}{\omega_n^2}
\sum_{\mu \nu}
\left .
\frac{\partial^2 F \left ( \textbf{R},\eta, 0 \right )}
{\partial \tau_{n} \partial \eta_{\mu \nu}}
\right |_{\textbf{R}_0,\eta_0}
\eta_{\mu \nu}^{\mathcal{E}_{\gamma}}.
\end{equation}
If we insert this relation into Eq. (\ref{eq_unclamp_chidc}) and use the
transformation Eq. (\ref{eq_transfo}) we obtain the following expression
of the unclamped EO tensor in the principal axes
\begin{eqnarray} \label{eq_unclamp_1}
r_{ij \gamma}^{\sigma}
= &&
\left .
\frac{-8 \pi}{n_i^2 n_j^2}
\chi_{ijl}^{(2)}
\right |_{l = \gamma}
-
\frac{4 \pi}{n_i^2 n_j^2 \sqrt{\Omega}}
\sum_m
\frac{\alpha_{ij}^m p_{m,\gamma}}{\omega_m^{2}} \nonumber \\
& & -
\frac{4 \pi}{n_i^2 n_j^2}
\sum_{\mu, \nu}
\left [
\left .
\frac{\partial \chi_{ij}^{(1)}
\left ( \textbf{R}, \eta, \bm{\mathcal{E}} \right )}
{\partial \eta_{\mu \nu}}
\right |_{\bm{R}_0, \eta_0, \mathcal{E}=0}
-
\frac{1}{\sqrt{\Omega}}
\sum_m
\frac{\alpha_{ij}^m}{\omega_m^2}
\left .
\frac{\partial^2 F \left ( \textbf{R},\eta, 0 \right )}
{\partial \tau_{m} \partial \eta_{\mu \nu}}
\right |_{\textbf{R}_0,\eta_0, \mathcal{E}=0}
\right ]
\eta_{\mu \nu}^{\mathcal{E}_{\gamma}}
\end{eqnarray}
The sum of the first and second term of the right-hand side of Eq.
(\ref{eq_unclamp_1}) is equal to the clamped EO coefficient
$r_{ij \gamma}^{\eta}$.
The product of the conversion factor
times the bracket in the third term of Eq. (\ref{eq_unclamp_1})
is equal to the elasto-optic coefficient
$p_{ij \mu \nu}$ [Eq. (\ref{eq_elastoopt})].
Finally, by definition of the converse piezoelectric effect,
$\eta_{\mu \nu}^{\mathcal{E}_{\gamma}}$
is equal to the piezoelectric strain coefficient
$d_{\gamma \mu \nu}$.
We obtain thus the following expression of the unclamped EO coefficients
that is equal to the one derived in Sec. \ref{sec_app_mac} from pure
macroscopic arguments
\begin{equation}
r_{ij \gamma}^{\sigma} = r_{ij \gamma}^{\eta} +
\sum_{\mu,\nu = 1}^{3} p_{ij \mu \nu} d_{\gamma \mu \nu}.
\end{equation}
It is worth noticing that the so-called piezoelectric contribution does
not only take into account the change of the linear optical
susceptibility with strain (third term of the right hand side of Eq.
(\ref{eq_unclamp_chidc})) but also includes the modification of the
ionic contribution, with respect to the clamped case, that is associated
to the modification of the ionic relaxation induced by the strain.


\begin{thebibliography}{}

\bibitem{hk}
P. Hohenberg et W. Kohn,
\newblock  Phys. Rev. {\bf 136}, B864 (1964).

\bibitem{ks}
W. Kohn et L. J. Sham,
\newblock  Phys. Rev. {\bf 140}, A1133 (1965).

\bibitem{rmp73_515}
S. Baroni, S. de Gironcoli, A. Dal Corso and P. Giannozzi,
\newblock Rev. Mod. Phys. {\bf 73}, 515 (2001).

\bibitem{irpc16_389}
B. Kirtman and B. Champagne,
\newblock International Reviews in Physical Chemistry {\bf 16}, 389 (1997).

\bibitem{champagne}
B. Champagne and B. Kirtman,
\newblock {\it Handbook of Advanced Electronic and Photonic Materials
and Devices}, edited by H. S. Nalwa, (Academic Press, 2001), Vol. 9,
p. 63.

\bibitem{ssc91_813}
A. Debernardi and S. Baroni,
\newblock Solid State Communications {\bf 91}, 813 (1994).

\bibitem{prl75_1819}
A. Debernardi, S. Baroni and E. Molinari,
\newblock Phys. Rev. Lett. {\bf 75}, 1819 (1995).

\bibitem{prb57_12847}
A. Debernardi,
\newblock Phys. Rev. B {\bf 57}, 12847 (1998).

\bibitem{prb53_15638}
A. Dal Corso, F. Mauri and A. Rubio,
\newblock Phys. Rev. B {\bf 53}, 15638 (1996).

\bibitem{prb66_100301}
G. Deinzer and D. Strauch,
\newblock Phys. Rev. B {\bf 66}, 100301 (2002).

\bibitem{prb67_144304}
G. Deinzer, G. Birner and D. Strauch,
\newblock Phys. Rev. B {\bf 67}, 144304 (2003).

\bibitem{prb69_14304}
G. Deinzer, M. Schmitt, A. P. Mayer, and D. Strauch,
\newblock Phys. Rev. B {\bf 69}, 14304 (2004).

\bibitem{prb69_45205}
G. Deinzer and D. Strauch,
\newblock Phys. Rev. B {\bf 69}, 45205 (2004).

\bibitem{prb63_155107}
R. W. Nunes and X. Gonze,
\newblock Phys. Rev. B {\bf 63}, 155107 (2001).

\bibitem{prl73_712}
R. W. Nunes and D. Vanderbilt,
\newblock Phys. Rev. Lett. {\bf 73}, 712 (1994).

\bibitem{abinit}
X. Gonze, J.-M. Beuken, R. Caracas, F. Detraux,
M. Fuchs, G.-M. Rignanese, L. Sindic, M. Verstraete,
G. Zerah, F. Jollet, M. Torrent, A. Roy, M. Mikami,
Ph. Ghosez, J.-Y. Raty and D.C. Allan,
\newblock Computational Materials Science {\bf 25}, 478 (2002).
[URL www.abinit.org]
 
\bibitem{prb39_13120}
X. Gonze and J.-P. Vigneron,
\newblock Phys. Rev. B {\bf 39}, 13120 (1989).

\bibitem{pra52_1086}
X. Gonze,
\newblock Phys. Rev. A {\bf 52}, 1086 (1995).

\bibitem{pra52_1096}
X. Gonze,
\newblock Phys. Rev. A {\bf 52}, 1096 (1995).

\bibitem{xggrad}
X. Gonze,
\newblock Phys. Rev. B {\bf 55}, 10337 (1997).

\bibitem{xgcl}
X. Gonze and C. Lee,
\newblock Phys. Rev. B {\bf 55}, 10355 (1997).

\bibitem{prb43_7231}
P. Giannozzi, S. de Gironcoli, P. Pavone and S. Baroni,
\newblock Phys. Rev. B {\bf 43}, 7231 (1991).


\bibitem{prb47_1651}
R. D. King-Smith and D. Vanderbilt,
\newblock Phys. Rev. B {\bf 47}, 1651 (1993).

\bibitem{marzari_wannier}
N. Marzari and D. Vanderbilt
\newblock Phys. Rev. B {\bf 56}, 12847 (1997).

\bibitem{prb48_4442}
D. Vanderbilt and R. D. King-Smith,
\newblock Phys. Rev. B {\bf 48}, 4442 (1993).




\bibitem{prb50_5756}
A. Dal Corso and F. Mauri,
\newblock Phys. Rev. B {\bf 50}, 5756 (1994).


\bibitem{prb53_10751}
J. L. P. Hughes and J. E. Sipe,
\newblock Phys. Rev. B {\bf 53}, 10751 (1996).

\bibitem{pr126_1977}
D. A. Kleinman,
\newblock Phys. Rev. {\bf 126}, 1977 (1962).

\bibitem{prl66_41}
Z. H. Levine and D. C. Allan,
\newblock Phys. Rev. Lett. {\bf 66}, 41 (1991).

\bibitem{prb44_12781}
Z. H. Levine and D. C. Allan,
\newblock Phys. Rev. B {\bf 44}, 12781 (1991).

\bibitem{prb49_4532}
Z. H. Levine,
\newblock Phys. Rev. B {\bf 49}, 4532 (1994).

\bibitem{prb47_9464}
M.Z. Huang, W.Y. Ching
\newblock Phys. Rev. B {\bf 47}, 9464 (1993).

\bibitem{prb67_165332}
S. Sharma, J. K. Dewhurst, C. Ambrosch-Draxl
\newblock Phys. Rev. B {\bf 67}, 165332 (2003).

\bibitem{cardona}
{\it Light Scattering in Solids II},
\newblock edited by M. Cardona and G. G\"untherodt
(Springer-Verlag, Berlin, 1982).

\bibitem{venkataraman}
{\it Dynamics of Perfect Crystals},
\newblock G. Venkataraman, L. A. Feldkamp and V. C. Sahni
(MIT Press, 1975).

\bibitem{prb1_3494}
W. D. Johnston Jr.,
\newblock Phys. Rev. B {\bf 1}, 3494 (1970).

\bibitem{prb33_5969}
S. Baroni and R. Resta,
\newblock Phys. Rev. B {\bf 33}, 5969 (1986).

\bibitem{prb63_94305}
P. Umari, A. Pasquarello and A. Dal Corso,
\newblock Phys. Rev. B {\bf 63}, 94305 (2001).

\bibitem{prl90_27401}
P. Umari, X. Gonze and A. Pasquarello,
\newblock Phys. Rev. Lett. {\bf 90}, 27401 (2003).

\bibitem{prl90_36401}
M. Lazzeri and F. Mauri,
\newblock Phys. Rev. Lett. {\bf 90}, 36401 (2003).

\bibitem{prb68_161101}
M. Lazzeri and F. Mauri,
\newblock Phys. Rev. B {\bf 68}, 161101 (2003).

\bibitem{prb1_910}
R. M. Pick, M. H. Cohen and R. M. Martin,
\newblock Phys. Rev. B {\bf 1}, 910 (1970).

\bibitem{sanchez}
F. Sanchez,
\newblock {\it Optique non-lin\'eaire} (Ellipses, 1999).

\bibitem{principal_axes}
In some cases, the electric field can induce a rotation of the principal
axes. Eq. \ref{eq_eo_el1} is expressed in the principal 
axes of the crystal at zero field.

\bibitem{pr160_519}
I. P. Kaminow and W. D. Johnston Jr.,
\newblock Phys. Rev. {\bf 160}, 519 (1967);
{\bf 178} 1528(E) (1969).

\bibitem{veithen_unpublished}
M. Veithen, X. Gonze and Ph. Ghosez,
\newblock unpublished.

\bibitem{asss3_264}
S. H. Wemple and D. DiDomenico Jr.,
\newblock in {\it Applied Solid State Science},
ed. R. Wolfe (Academic, N.Y., 1972).



\bibitem{bbbcg}
M. C. Payne, M. P. Teter, D. C. Allan, T. A. Arias et J. D. Joannopoulos,
\newblock Rev. Mod. Phys {\bf 64}, 1045 (1992).

\bibitem{prb45_13244}
J. P. Perdew and Y. Wang,
\newblock Phys. Rev. B {\bf 45}, 13244 (1992).

\bibitem{prb54_1703}
S. Goedecker, M. Teter and J. Hutter,
\newblock Phys. Rev. B {\bf 54}, 1703 (1996).

\bibitem{prb43_1993}
N. Troullier and J. L. Martins,
\newblock Phys. Rev. B {\bf 43}, 1993 (1991).

\bibitem{prb48_5031}
M. P. Teter,
\newblock Phys. Rev. B {\bf 48}, 5031 (1993).

\bibitem{nature358_137}
R. E. Cohen,
\newblock Nature {\bf 358}, 137 (1992).

\bibitem{prl74_4035}
X. Gonze, Ph. Ghosez and R. W. Godby,
\newblock Phys. Rev. Lett. {\bf 74}, 4035 (1995).

\bibitem{epl33_713}
Ph. Ghosez, X. Gonze and J. -P. Michenaud,
\newblock Europhysics Letters {\bf 33}, 713 (1996).

\bibitem{prl63_1719}
Z. H. Levine and D. C. Allan,
\newblock Phys. Rev. Lett. {\bf 63}, 1719 (1989).

\bibitem{prl89_117602}
I. Souza, J. \'I\~niguez and D. Vanderbilt,
\newblock Phys. Rev. Lett. {\bf 89}, 117602 (2002).

\bibitem{prb54_8540}
W. G. Aulbur, Lars J\"onsson and J. W. Wilkins,
\newblock Phys. Rev. B {\bf 54}, 8540 (1996).

\bibitem{ssc48_301}
J. Wagner and M. Cardona,
\newblock Solid State Communications {\bf 48}, 301 (1983).

\bibitem{jpcs61_147}
D. Vanderbilt,
\newblock J. Phys. Chem. Solids {\bf 61}, 147 (2000).

\bibitem{prb65_214302}
M. Veithen and Ph. Ghosez,
\newblock Phys. Rev. B {\bf 65}, 214302 (2002).

\bibitem{prb48_10160}
C. M. Foster, Z. Li, M. Grimsditch, S. -K. Chan, and D. J. Lam,
\newblock Phys. Rev. B {\bf 48}, 10160 (1993).

\bibitem{pr188_1209}
W. D. Johnston Jr. and I. P. Kaminow,
\newblock Phys. Rev. {\bf 188}, 1209 (1969).

\bibitem{pma45_239}
J. M. Calleja, H. Vogt and M. Cardona,
\newblock Philosophical Magazine A {\bf 45}, 239 (1982).

\bibitem{jap78_2651}
P. Bernasconi, M. Zhonik and P. G\"unter,
\newblock J. Appl. Phys. {\bf 78}, 2651 (1995).

\bibitem{nye}
J. F. Nye,
\newblock {\it Physical properties of crystals},
(Oxford University Press, 1985).

\bibitem{landlif}
L. D. Landau and E. M. Lifshitz,
\newblock {\it Electrodynamics of continuous media},
Volume 8 of Course of Theoretical Physics
(Pergamon Press, 1960).

\bibitem{umesh}
U. V. Waghmare,
\newblock unpublished.

\bibitem{dc_phd}
A. Dal Corso,
\newblock {\it Density-functional theory beyond the pseudopotential
local density approach: a few cases studies}, PhD thesis, SISSA
(1993).



\bibitem{prb63_115118}
F. Detraux and X. Gonze,
\newblock Phys. Rev. B {\bf 63}, 115118 (2001).

\bibitem{mrs718_323}
K. M. Rabe,
\newblock Materials Research Society Symposium Proceedings {\bf 718},
323 (2002).

\bibitem{rauber}
A. R\"auber,
\newblock Current Topics in Materials Science, vol. 1,
ed. E. Kaldis (North-Holland, 1978), p. 481.

\bibitem{jap84_2251}
F. Abdi, M. Aillerie, P. Bourson, M. D. Fontana and K. Polgar,
\newblock J. Appl. Phys. {\bf 84}, 2251 (1998).

\bibitem{handbook_laser}
{\it Handbook of Laser Science and Technology}, Optical Materials:
Part 2 Vol. IV, ed. M. J. Weber
(CRC, Boca Raton, FL, 1986).

\bibitem{prb50_5941}
M. Zgonik {\it et al.},
\newblock Phys. Rev. B {\bf 50}, 5941 (1994).

\end{thebibliography}
\end{document}